\documentclass[aps,preprint,superscriptaddress,prb]{revtex4-2}
\usepackage{amsmath}
\usepackage{amssymb}
\usepackage{graphicx}
\usepackage[english]{babel}
\usepackage{bm}
\usepackage{mathrsfs}
\usepackage{hyperref}
\hypersetup{
    colorlinks=true,
    linkcolor=blue,
    citecolor=blue,      
    urlcolor=blue,
}

\newcommand{\op}[2][b]{\hat{#1}_{\mathbf{#2}}}
\newcommand{\opc}[2][b]{\hat{#1}^{\dagger}_{\mathbf{#2}}}

\newcommand{\braket}[1]{\left\langle #1 \right\rangle}
\newcommand{\brakett}[3]{\left\langle #1 \left| #2 \right| #3 \right\rangle}
\newcommand{\omegaLO}{\omega_{\text{LO}}}
\newcommand{\figref}[1]{Fig. \ref{#1}}
\newcommand{\secref}[1]{Sec. \ref{#1}}
\newcommand{\tabref}[1]{Table \ref{#1}}
\newcommand{\tildeff}{\mathop{}\!\mathrm{d}}

\begin{document}

\title{Optical conductivity of an anharmonic large polaron gas at weak coupling}

\author{Matthew Houtput}
\affiliation{Theory of Quantum Systems and Complex Systems, Universiteit Antwerpen, B-2000 Antwerpen, Belgium}

\author{Jacques Tempere}
\affiliation{Theory of Quantum Systems and Complex Systems, Universiteit Antwerpen, B-2000 Antwerpen, Belgium}

\begin{abstract}
In a polar solid, electrons or other charge carriers can interact with the phonons of the ionic lattice, leading to the formation of polaron quasiparticles. The optical conductivity and optical absorption spectrum of a material are affected by this electron-phonon coupling, most notably leading to an absorption peak in the mid-infrared region. Recently, a model Hamiltonian for anharmonic electron-phonon coupling was derived [M. Houtput and J. Tempere, \href{https://doi.org/10.1103/PhysRevB.103.184306}{Phys. Rev. B \textbf{103}, 184306 (2021)}], that includes both the conventional Fr\"ohlich interaction as well an interaction where an electron interacts with two phonons simultaneously. In this article, we calculate and investigate the optical conductivity of the anharmonic large polaron gas, and show that an additional characteristic absorption peak appears due to this 1-electron-2-phonon interaction.

We calculate a semi-analytical expression for the optical conductivity $\sigma(\omega)$ at finite temperatures and weak coupling using the Kubo formula. The electronic and phononic contributions can be split and treated separately, such that the many-body effects of the electron gas may be taken into account through the well-known dynamical structure factor $S(\mathbf{k},\omega)$. From the resulting optical conductivity, we calculate the polaron effective mass, an estimate for the electron-phonon scattering times, and the optical absorption spectrum of the anharmonic polaron gas. It is shown that the effects are negligible for four common III-V semiconductors (BN, AlN, BP, AlP) in the zincblende structure, which justifies the commonly used harmonic approximation in these materials. We show that alongside the well-known polaron absorption peak at the phonon energy $\hbar \omegaLO$, the 1-electron-2-phonon interaction leads to an additional absorption peak at $2 \hbar \omegaLO$. We propose this absorption peak as an experimentally measurable indicator for nonnegligible 1-electron-2-phonon interaction in a material, since the height of this peak is proportional to the strength of this anharmonic interaction.

\end{abstract}

\maketitle

\section{Introduction} \label{sec:Introduction}
A free electron moving through a lattice of ions can interact with the phonons of the lattice, for example by Coulomb interaction with these ions. This electron-phonon interaction will cause the electron to become dressed by the phonons, leading to the polaron quasiparticle. The polaron problem is nearly a century old \cite{Landau1933, Landau1948} and has been extensively studied ever since its prediction. Since it is one of the simplest models of an impurity interacting with a bosonic field, many analogies of the polaron exist, including spin polarons \cite{Nagaev1974}, exciton polarons \cite{Verzelen2002}, ripplopolarons \cite{Tempere2003}, magnetic polarons \cite{Koepsell2019}, and the Bose \cite{Jorgensen2016,Shchadilova2016}
and Fermi \cite{Schirotzek2009} polaron in ultracold gases.

Many of the properties of a polaron are different from that of a free electron. It is well-known that due to the electron-phonon interaction, the polaron has a lower ground state energy and a higher effective mass \cite{Lee1953, Frohlich1954, Feynman1955, Alexandrov2010}. Quite importantly, the response properties of the polaron are also different. Collisions of the polaron with phonons cause the material to have a finite DC conductivity \cite{Drude1900_1, Drude1900_2, Kadanoff1963}. Additionally, in the weak electron-phonon coupling limit, the optical absorption spectrum of a polaron has an additional absorption peak in the mid-infrared region near the phonon energy, caused by the elementary phonon emission process \cite{Gurevich1962, Tempere2001, Finkenrath1969, VanMechelen2008}. Both of these effects are described by the optical conductivity $\sigma(\omega)$, which describes the response of the polaron with respect to an electric field. It can be calculated using several different methods \cite{Feynman1962, Devreese1972, Peeters1983, Mishchenko2003}; calculations based on the Kubo formula are among the most popular methods when calculating the conductivity in the weak-coupling limit \cite{Kubo1957_1, Kubo1957_2, Mahan2000, Tempere2001}.

\begin{figure}
\centering
\includegraphics[width=8.6cm]{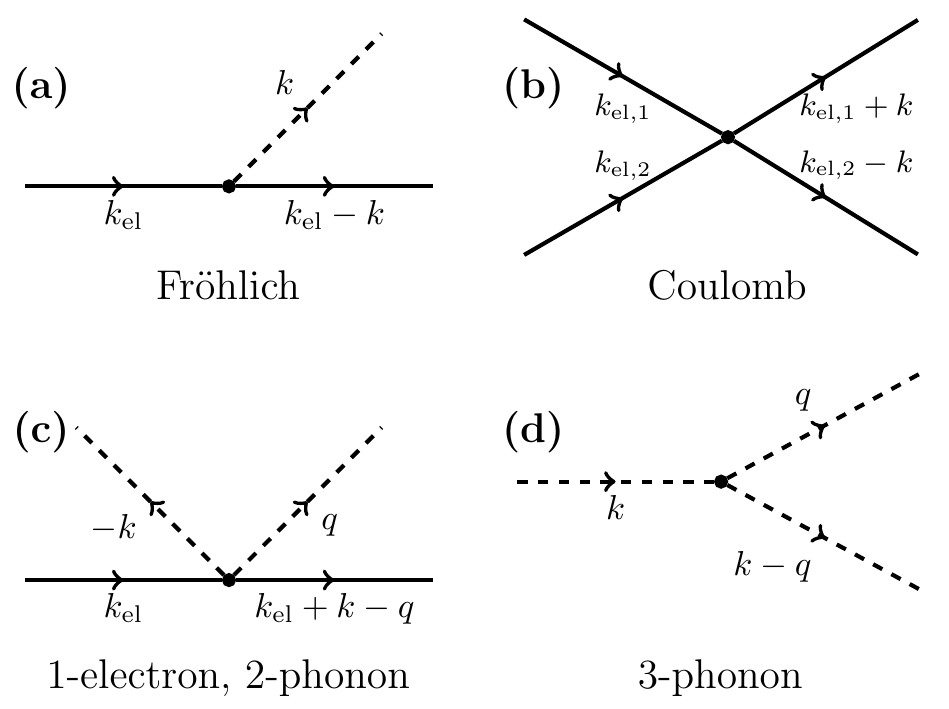}
\caption{\label{fig:HamRepresentation} Feynman diagrams of the different interactions between electrons and phonons that are considered in this article. Solid lines represent electrons, and dashed lines represent phonons. The Fr\"ohlich and Coulomb interactions are considered in the usual large polaron treatment; the novelty of this article is the inclusion of interactions \textbf{(c)} and \textbf{(d)}. In the present analysis text, the dashed lines represent LO phonons only.}
\end{figure}

If the material under consideration is a polar semiconductor, and the electron wavefunction is sufficiently large (a so-called ``large'' polaron) the electron-phonon coupling is usually well described by the Fr\"ohlich Hamiltonian \cite{Frohlich1954}. In this Hamiltonian, it is assumed that the electron-phonon coupling is linear, as in \figref{fig:HamRepresentation}a, and the electron only interacts with longitudinal optical (LO) phonons. In recent years, however, it has been shown that in some materials, other interactions play a nonnegligible role. Recent work in SrTiO$_3$ \cite{Gastiasoro2020} has shown the importance of an interaction term of the form shown in \figref{fig:HamRepresentation}c, where an electron interacts simultaneously with two transverse optical (TO) phonons \cite{Ngai1974}. This interaction has been proposed as a mechanism for superconductivity in SrTiO$_3$ \cite{VanDerMarel2019, Feigelman2021}, and has been used to explain the anomalous $T^2$-behavior of the resistivity at low temperatures \cite{Kumar2021}. In several hydrogen-rich materials under extreme pressures \cite{Drozdov2015, Somayazulu2019, Errea2015} and potentially also metallic hydrogen \cite{Ashcroft1968, Dias2017, Loubeyre2020}, phonon-mediated superconductivity is possible at temperatures very close to room temperature. However, since hydrogen-rich materials are strongly anharmonic, interactions between the phonons like in \figref{fig:HamRepresentation}d and potentially also interactions like \figref{fig:HamRepresentation}c must be taken into account. Finally, similar 1-electron-2-phonon interactions and 3-phonon interactions are also present in the ultracold Bose polaron, where they lead to a significant change in the energy \cite{Rath2013, Ichmoukhamedov2019}.

In a recent paper \cite{Houtput2021}, analytical expressions were derived for the interaction strengths of the 1-electron-2-LO-phonon interaction (\figref{fig:HamRepresentation}c) and the 3-LO-phonon interaction (\figref{fig:HamRepresentation}d), suitable under the same conditions that are used for the Fr\"ohlich Hamiltonian. The goal of this article is to calculate the optical conductivity $\sigma(\omega)$ of a polaron where these anharmonic interaction terms are also taken into account. We will calculate the conductivity in the weak-coupling limit using the method proposed in \cite{Tempere2001}, which also yields the conductivity of a many-polaron gas without much additional effort. Since the interaction strengths are known analytically just as for the Fr\"ohlich Hamiltonian \cite{Frohlich1954, Houtput2021}, the derivation may also proceed analytically.

The structure of this paper is as follows. In \secref{sec:Theory}, the model Hamiltonian is outlined, and an expression for the conductivity in terms of the dynamical structure factor of the electron gas is obtained. In \secref{sec:Results}, the various limits of this expression are examined in detail, and the effect of the anharmonic interaction terms is investigated. We conclude in \secref{sec:Conclusion}.

\section{Theory} \label{sec:Theory}
\subsection{Extended Fr\"ohlich Hamiltonian}
The Hamiltonian that will be studied in this article is an extension of the well-known Fr\"ohlich Hamiltonian \cite{Frohlich1954}, which is derived for a polar cubic lattice with two atoms in the primitive unit cell. It must additionally be assumed that the crystal has no inversion symmetry \cite{Houtput2021}. Under all these assumptions, the anharmonic polaron Hamiltonian including 3-phonon and 1-electron-2-phonon interaction is:
\begin{equation} \label{HamTot}
\hat{H} = \hat{H}_{\text{el}} + \hat{H}_{\text{ph}} + \hat{H}_{\text{el-ph}},
\end{equation}
where the three terms in the Hamiltonian represent the electron Hamiltonian, the phonon Hamiltonian, and the electron-phonon interaction \cite{Mahan2000, Houtput2021}:
\begin{align}
\hat{H}_{\text{el}} & = \sum_{\mathbf{k}} \varepsilon_{\mathbf{k}} \opc[c]{k} \op[c]{k} +  \frac{1}{2} \sum_{\mathbf{k},\mathbf{k}'} \sum_{\mathbf{q} \neq \mathbf{0}} \mathcal{V}^{(C)}_{\mathbf{q}} \hat{c}^{\dagger}_{\mathbf{k}+\mathbf{q}} \hat{c}^{\dagger}_{\mathbf{k}'-\mathbf{q}} \hat{c}_{\mathbf{k}'} \hat{c}_{\mathbf{k}},  \label{HamEl} \\
\hat{H}_{\text{ph}} & = \sum_{\mathbf{q}} \hbar \omega_{\mathbf{q}} \left( \opc{q} \op{q} + \frac{1}{2} \right) + \frac{1}{6}\sum_{\mathbf{q} \neq \mathbf{q}' \neq \mathbf{0}} \mathcal{V}^{(0)}_{\mathbf{q},\mathbf{q}'} \left( \opc{-q} + \op{q} \right)\left( \opc{q-q'} + \op{-q+q'} \right)\left( \opc{q'} + \op{-q'} \right), \label{HamPh} \\
\hat{H}_{\text{el-ph}} & = \sum_{\mathbf{q} \neq \mathbf{0}} \mathcal{V}^{(F)}_{\mathbf{q}} \left( \opc{q} + \op{-q} \right) \hat{\rho}_{-\mathbf{q}} + \frac{1}{2} \sum_{\mathbf{q} \neq \mathbf{q}' \neq \mathbf{0}} \mathcal{V}^{(1)}_{\mathbf{q},\mathbf{q}'} \left( \opc{-q} + \op{q} \right)\left( \opc{q'} + \op{-q'} \right) \hat{\rho}_{\mathbf{q}-\mathbf{q}'}. \label{HamElPh}
\end{align}
The sums over $\mathbf{q}$ and $\mathbf{q}'$ in \eqref{HamEl}-\eqref{HamElPh} exclude the cases where $\mathbf{q} = \mathbf{0}, \mathbf{q}' = \mathbf{0}$ and $\mathbf{q} = \mathbf{q}'$. It is usually more convenient to take this into account by requiring that $\mathcal{V}^{(F)}_{\mathbf{0}} = 0$, $\mathcal{V}^{(n)}_{\mathbf{0},\mathbf{q}} = 0$, $\mathcal{V}^{(n)}_{\mathbf{q},\mathbf{q}} = 0$, and so on.
$\opc{q}, \op{q}$ and $\opc[c]{k}, \op[c]{k}$ represent the creation and annihilation operators of the LO phonons and the electrons, respectively, and the electron density operator $\hat{\rho}_{\mathbf{k}}$ is given by:
\begin{equation}
\hat{\rho}_{\mathbf{k}} = \sum_{\mathbf{k}'} \opc[c]{k+k'} \op[c]{k'}.
\end{equation}
As in the Fr\"ohlich model \cite{Frohlich1954}, it is assumed that the electrons occupy a single parabolic band described by a band mass $m_b$, and that the electrons only interact with a single LO phonon branch which has a constant frequency $\omegaLO$:
\begin{align}
\varepsilon_{\mathbf{k}} & = \frac{\hbar^2 k^2}{2m_b}, \label{epsilonK} \\
\omega_{\mathbf{q}} & = \omegaLO.
\end{align}
The electron Hamiltonian \eqref{HamEl} contains the Coulomb interaction of \figref{fig:HamRepresentation}b, the phonon Hamiltonian \eqref{HamPh} contains the 3-phonon interaction of \figref{fig:HamRepresentation}d, and both the Fr\"ohlich interaction of \figref{fig:HamRepresentation}a and the 1-electron-2-phonon interaction of \figref{fig:HamRepresentation}c are included in the electron-phonon interaction term \eqref{HamElPh}. In the large polaron limit, the interaction strengths have the following analytical expressions \cite{Houtput2021}:
\begin{align}
\mathcal{V}^{(C)}_{\mathbf{q}} & = \frac{e^2}{V \varepsilon_{\text{vac}} \varepsilon_{\infty}} \frac{1}{|\mathbf{q}|^2},
\label{VCoulomb} \\
\mathcal{V}^{(F)}_{\mathbf{q}} & = \hbar \omegaLO \sqrt{\frac{4\pi\alpha}{V} } \left(\frac{\hbar}{2m_b \omegaLO} \right)^{\frac{1}{4}} \frac{1}{|\mathbf{q}|}, \label{VFrohlich} \\
\mathcal{V}^{(0)}_{\mathbf{q},\mathbf{q}'} & = -i \hbar \omegaLO \frac{\mathcal{T}_0}{\sqrt{V}} \left(\frac{\hbar}{2m_b \omegaLO} \right)^{\frac{3}{4}} |\varepsilon_{ijl}| \frac{q_i (q_j-q'_j) q'_j}{|\mathbf{q}||\mathbf{q}-\mathbf{q}'||\mathbf{q}'|}, \label{VAnhar0} \\
\mathcal{V}^{(1)}_{\mathbf{q},\mathbf{q}'} & = -i \hbar \omegaLO \frac{\sqrt{4\pi \alpha} \mathcal{T}_1}{V} \frac{\hbar}{2m_b \omegaLO} |\varepsilon_{ijl}| \frac{q_i (q_j-q'_j) q'_j}{|\mathbf{q}||\mathbf{q}-\mathbf{q}'|^2|\mathbf{q}|}. \label{VAnhar1}
\end{align}
In these expressions $\varepsilon_{\text{vac}}$ is the vacuum permittivity in SI units, $|\varepsilon_{ijl}|$ is the absolute value of the Levi-Civita tensor, $\alpha$ is the Fr\"ohlich electron-phonon coupling constant \cite{Frohlich1954, Mahan2000}, and $\mathcal{T}_0$ and $\mathcal{T}_1$ are dimensionless material parameters which respectively characterize the strength of the 3-phonon and 1-electron-2-phonon interaction. All results will be plotted in terms of the coupling constants $\alpha$, $\mathcal{T}_0$, and $\mathcal{T}_1$, which are assumed to be known. We present values for some of these parameters in \secref{sec:MaterialParameters}.

A remark must be made on the applicability of the Hamiltonian \eqref{HamTot}-\eqref{HamElPh}. While many assumptions were made in its derivation, there is a broad class of III-V semiconductors in the zincblende structure which satisfy all these assumptions \cite{Houtput2021}. To have a concrete example in mind during the calculations, a semiconductor with the zincblende structure (such as AlN or GaAs) is assumed for the remainder of the article.

\subsection{Memory function formalism}
To calculate the conductivity of a gas of $N$ polarons in the weak-coupling limit, we follow the method used in \cite{Tempere2001} which is based on the Kubo-Greenwood formula \cite{Mahan2000}. It relates the conductivity to a momentum-momentum correlation function:
\begin{equation} \label{condPP}
\sigma(\omega) = \lim_{\delta \rightarrow 0^+} \left( i \frac{ne^2}{m_b (\omega+i\delta)}  + \frac{n e^2}{ N m_b^2 \hbar (\omega+i\delta)} \int_{0}^{+\infty} \left\langle \left[ \hat{P}_{x}(t), \hat{P}_{x}(0) \right] \right\rangle e^{i (\omega+i\delta) t} \tildeff t \right),
\end{equation}
where $\hat{\mathbf{P}} = \sum_{i=1}^N \hat{\mathbf{p}}_{\text{el},i}$ is the total electron momentum operator defined in the Heisenberg picture, and $n := N/V$ is the electron density. Note that the conductivity can be represented by a scalar, since cubic symmetry is assumed. When calculating the conductivity of polarons, it is customary to write this function in the following form \cite{Peeters1983, Klimin2020}:
\begin{equation} \label{condMemory}
\sigma(\omega) = \lim_{\delta \rightarrow 0^+} i \frac{n e^2}{m_b} \frac{1}{\omega +i \delta - \Sigma(\omega)},
\end{equation}
where the memory function $\Sigma(\omega)$ is defined as:
\begin{align}
\Sigma(\omega) & = \lim_{\delta \rightarrow 0^+} \frac{\Sigma_0(\omega+i\delta)}{1 +  \frac{\Sigma_0(\omega+i\delta)}{\omega +i \delta}} = \frac{\Sigma_0(\omega)}{1 +  \frac{\Sigma_0(\omega)-\Sigma_0(0)}{\omega}}, \label{MemoryFullDef} \\
\Sigma_0(\omega) & := \lim_{\delta \rightarrow 0^+} \frac{(\omega+i\delta)}{iN m_b \hbar} \int_{0}^{+\infty} e^{i(\omega+i\delta)t} \left\langle \left[ \hat{P}_{x}(t), \hat{P}_{x}(0) \right] \right\rangle  \tildeff t. \label{Sigma0Def}
\end{align}
In practice, all the information of the conductivity is now contained in the simpler function $\Sigma_0(\omega)$, which is written in terms of a retarded momentum-momentum Green's function. In the weak-coupling limit, $\Sigma_0(\omega)$ can be calculated using a Green's function diagrammatic expansion \cite{Mahan2000}. Once $\Sigma_0(\omega)$ is known, the conductivity can be calculated using the algebraic formulas \eqref{condMemory}-\eqref{MemoryFullDef}.

In \cite{Tempere2001, Alexandrov2010}, it is shown that $\Sigma_0(\omega)$ can also be written in terms of a force-force correlation function by applying two partial integrations to equation \eqref{Sigma0Def}:
\begin{equation} \label{Sigma0FF}
\Sigma_0(\omega) := \lim_{\delta \rightarrow 0^+} \frac{1}{i N m_b \hbar (\omega+i\delta)} \int_{0}^{+\infty} e^{-\delta t} \left(e^{i \omega t}-1 \right) \left\langle \left[ \hat{F}_{x}(t), \hat{F}_{x}(0) \right] \right\rangle  \tildeff t,
\end{equation}
where the force operator $\hat{\mathbf{F}}$ is defined as:
\begin{equation} \label{ForceDef}
\hat{\mathbf{F}}(t) := \frac{d\hat{\mathbf{P}}}{dt} = \frac{i}{\hbar} [\hat{H},\hat{\mathbf{P}}(t)].
\end{equation}
The force operator can be calculated exactly by plugging in the Hamiltonian \eqref{HamTot}-\eqref{HamElPh} into equation \eqref{ForceDef}. Since the electron Hamiltonian $\hat{H}_{\text{el}}$ conserves the total electron momentum $\hat{\mathbf{P}}(t)$, and all phonon operators commute with $\hat{\mathbf{P}}(t)$, it holds that $[\hat{H}_{\text{el}}, \hat{\mathbf{P}}]=[\hat{H}_{\text{ph}}, \hat{\mathbf{P}}] = \mathbf{0}$, so that only the interaction Hamiltonian $\hat{H}_{\text{int}}$ \eqref{HamElPh} contributes to the force operator. Making use of the identity $[\hat{\rho}_{\mathbf{k}}, \hat{\mathbf{P}}] = -\hbar \mathbf{k} \hat{\rho}_{\mathbf{k}}$, the force operator can be written as follows:
\begin{equation} \label{FExpr}
\hat{\mathbf{F}}(t) = i \sum_{\mathbf{k}} \mathbf{k} \hat{\mathcal{F}}_{\mathbf{k}} \hat{\rho}_{-\mathbf{k}},
\end{equation}
where the auxiliary operator $\hat{\mathcal{F}}_{\mathbf{k}}$ is a bosonic operator related to the phonon operators:
\begin{equation} \label{FAuxDef}
\hat{\mathcal{F}}_{\mathbf{k}} := \mathcal{V}^{(F)}_{\mathbf{k}} \left( \opc{k} + \op{-k} \right) + \frac{1}{2} \sum_{\mathbf{q}} \mathcal{V}^{(1)}_{-\mathbf{k}+\mathbf{q},\mathbf{q}} \left( \opc{k-\mathbf{q}} + \op{-k+\mathbf{q}} \right) \left( \opc{q} + \op{-q} \right).
\end{equation}
Only the electron-phonon interaction terms \eqref{HamElPh} contribute to the force operator \eqref{FExpr}. This means the product $\hat{F}_x(t) \hat{F}_x(0)$ in equation \eqref{Sigma0FF} is proportional to $\alpha$: indeed, in expression \eqref{FAuxDef}, both $\mathcal{V}^{(F)}_{\mathbf{k}}$ and $\mathcal{V}^{(1)}_{-\mathbf{k}+\mathbf{q},\mathbf{q}}$ are proportional to $\sqrt{\alpha}$ (see expressions \eqref{VCoulomb}-\eqref{VAnhar1}). Since one factor $\alpha$ is factored out beforehand, this means $\Sigma_0(\omega)$ can be calculated exactly up to first order in $\alpha$. In particular, the expectation values with respect to the electron and phonon operators can be factorized, since for any electron operator $\hat{A}_{\text{el}}$ and any phonon operator $\hat{B}_{\text{ph}}$, it holds that:
\begin{equation} \label{ExpectationFactorize}
\braket{\hat{A}_{\text{el}} \hat{B}_{\text{ph}}} = \braket{\hat{A}_{\text{el}}}_0 \braket{\hat{B}_{\text{ph}}}_0 + O(\alpha),
\end{equation}
where $\braket{}_0$ indicates an expectation value with respect to $\hat{H}_{\text{el}}$ \eqref{HamEl} for the electron operators, and with respect to $\hat{H}_{\text{ph}}$ \eqref{HamPh} for the phonon operators. This factorization means the electron and phonon problems can be treated separately.

Using \eqref{FExpr} and \eqref{ExpectationFactorize}, the force-force correlation function in \eqref{Sigma0FF} can be written as:
\begin{equation} \label{Fcorrint1}
\left\langle \left[ \hat{F}_{x}(t), \hat{F}_{x}(0) \right] \right\rangle = \frac{1}{3} \sum_{\mathbf{k}} k^2 \text{Im}\left[ \left\langle \hat{\mathcal{F}}_{\mathbf{k}}(t) \hat{\mathcal{F}}_{-\mathbf{k}}(0) \right\rangle_0 \left \langle \hat{\rho}_{-\mathbf{k}}(t) \hat{\rho}_{\mathbf{k}}(0) \right\rangle_0 \right] + O(\alpha^2),
\end{equation}
where we used that $\left \langle \hat{\rho}_{-\mathbf{k}}(t) \hat{\rho}_{\mathbf{q}}(0) \right\rangle_0$ is zero unless $\mathbf{k} = \mathbf{q}$. This can be understood by noting that the density operator $\hat{\rho}_{\mathbf{k}}$ adds a momentum $\mathbf{k}$ to the electron system, which must be removed again to end up in the same state. Additionally, since the system is isotropic, $k_x^2$ was replaced by $(k_x^2 + k_y^2 + k_z^2)/3 = k^2/3$.

In order to calculate the expectation values appearing in \eqref{Fcorrint1}, we note that they can be related to more familiar quantities from the literature. In particular, the expectation value of the electron operators is the inverse Fourier transform of the dynamical structure factor $S(\mathbf{k},\omega)$ of the electron gas \cite{Mahan2000, Tempere2001}:
\begin{equation} \label{StructureFactorDef}
S(\mathbf{k},\omega) := \frac{1}{2\pi N} (1-e^{-\hbar \beta \omega}) \int_{-\infty}^{+\infty} \left \langle \hat{\rho}_{\mathbf{k}}(t) \hat{\rho}_{-\mathbf{k}}(0) \right\rangle_0 e^{i \omega t} \tildeff t.
\end{equation}
A similar quantity can be defined for the phonon operators, which we will refer to as the phonon spectral function $M(\mathbf{k},\omega)$:
\begin{equation} \label{PhononMatrixDef}
M(\mathbf{k},\omega) := \frac{1}{2\pi} (1-e^{-\hbar \beta \omega}) \int_{-\infty}^{+\infty} \left\langle \hat{\mathcal{F}}_{-\mathbf{k}}(t) \hat{\mathcal{F}}_{\mathbf{k}}(0) \right\rangle_0 e^{i \omega t} \tildeff t.
\end{equation}
The phonon spectral function $M(\mathbf{k},\omega)$ will be calculated in \secref{sec:PhononSpectralFunction}, and the possible models for the dynamical structure factor $S(\mathbf{k},\omega)$ will be discussed in \secref{sec:DynamicalStructureFactor}. First, the force-force correlation function \eqref{Fcorrint1} will be rewritten in terms of these two quantities.

Inverting the Fourier transforms in \eqref{StructureFactorDef}-\eqref{PhononMatrixDef} gives useful expressions for the expectation values appearing in \eqref{Fcorrint1}:
\begin{align}
 \left \langle \hat{\rho}_{-\mathbf{k}}(t) \hat{\rho}_{\mathbf{k}}(0) \right\rangle_0 = & N \int_{-\infty}^{+\infty} \frac{S(-\mathbf{k},\omega)}{1-e^{-\hbar \beta \omega}} e^{-i \omega t} \tildeff \omega \\
\left\langle \hat{\mathcal{F}}_{\mathbf{k}}(t) \hat{\mathcal{F}}_{-\mathbf{k}}(0) \right\rangle_0 = & \int_{-\infty}^{+\infty} \frac{M(\mathbf{k},\omega)}{1-e^{-\hbar \beta \omega}} e^{-i \omega t} \tildeff \omega
\end{align}
With these expressions, equation \eqref{Fcorrint1} for the force-force correlation function can be straightforwardly calculated. The result can be written as follows:
\begin{align}
& \left\langle \left[ \hat{F}_{x}(t), \hat{F}_{x}(0) \right] \right\rangle = \\
& -\frac{2Ni}{3} \sum_{\mathbf{k}} k^2 \int_0^{+\infty} \left\{ \int_{-\infty}^{+\infty} \left[1+n_B(\omega')+n_B(\omega-\omega')\right] S(\mathbf{k},\omega-\omega') M(\mathbf{k},\omega') d\omega' \right\} \sin(\omega t) \tildeff\omega,
\end{align}
where $n_B(\omega) = 1/(e^{\hbar \beta \omega}-1)$. Plugging this force-force correlation function back into \eqref{Sigma0FF}, another straightforward calculation shows that $\Sigma_0(\omega)$ can be written in terms of $S(\mathbf{k},\omega)$ and $M(\mathbf{k},\omega)$:
\begin{align}
\text{Re}\left[\Sigma_0(\omega)\right] = & \frac{2 \omega}{\pi} \mathcal{P} \int_{0}^{+\infty}  \frac{\text{Im}\left[\Sigma_0(\omega')\right]}{\omega'^2-\omega^2} 
\tildeff \omega', \label{SigmaReGen} \\
\text{Im}\left[\Sigma_0(\omega)\right] = & -\frac{\pi}{3m_b \hbar \omega} \sum_{\mathbf{k}} k^2 \int_{-\infty}^{+\infty} \left[1+n_B(\omega')+n_B(\omega-\omega')\right] S(\mathbf{k},\omega-\omega') M(\mathbf{k},\omega') d\omega' \nonumber \\
& \hspace{10pt} + O(\alpha^2).  \label{SigmaImGen}
\end{align}
Note that \eqref{SigmaReGen} is similar to the usual Kramers-Kronig relations, but is not exactly the form found in the literature \cite{Arfken2013} for a function $f(\omega)$ on the domain $\omega \in [0,+\infty[$. This is because the usual Kramers-Kronig relations are derived for a function satisfying $f(-\omega) = f^*(\omega)$, whereas equation \eqref{SigmaReGen} is for a function satisfying $\Sigma_0(-\omega) = -\Sigma_0^*(\omega)$. 

Equations \eqref{SigmaReGen}-\eqref{SigmaImGen} in combination with \eqref{condMemory}-\eqref{MemoryFullDef} allow for the calculation of the conductivity $\sigma(\omega)$ up to first order in $\alpha$ if the dynamical structure factor of the electron gas $S(\mathbf{k},\omega)$ and the phonon spectral function $M(\mathbf{k},\omega)$ given by \eqref{StructureFactorDef}-\eqref{PhononMatrixDef} are known. The dynamical structure factor of the electron gas is well-known in the literature, and is related to the dielectric function of the electron gas \cite{Lindhard1954, Devreese1980_1, Devreese1980_2, Mahan2000, Tempere2001}. The phonon spectral function $M(\mathbf{k},\omega)$ is the spectral function associated with the operator $\hat{\mathcal{F}}_{\mathbf{k}}$, and can be calculated analytically. Both of these quantities are discussed in the following sections.

\subsection{Calculation of the phonon spectral function \texorpdfstring{$M(\mathbf{k},\omega)$}{}} \label{sec:PhononSpectralFunction}
The phonon spectral function can be calculated using a Matsubara-Green diagrammatic expansion \cite{Mahan2000}. First, the definition \eqref{PhononMatrixDef} is rewritten in terms of a retarded Green's function:
\begin{align}
M(\mathbf{k},\omega) & = -\frac{1}{\pi} \text{Im} \left[ F_{\text{ret}}(\mathbf{k},\omega) \right], \label{Mretarded} \\
F_{\text{ret}}(\mathbf{k},\omega) & = -i \int_{0}^{+\infty} \left\langle \left[ \hat{\mathcal{F}}_{-\mathbf{k}}(t), \hat{\mathcal{F}}_{\mathbf{k}}(0) \right] \right\rangle e^{i \omega t} \tildeff t. \label{Fretarded}
\end{align}
It can straightforwardly be shown that the definition \eqref{Mretarded}-\eqref{Fretarded} is equivalent to the original definition \eqref{PhononMatrixDef}. It is rewritten using a retarded Green's function because a theorem by Matsubara \cite{Matsubara1955, Mahan2000} states that this retarded Green's function $F_{\text{ret}}(\mathbf{k},\omega)$ can be written as the analytic continuation of the Matsubara Green's function $\mathcal{F}(\mathbf{k},i \omega_n)$ \cite{Matsubara1955, Mahan2000}, which is a time-ordered Green's function in imaginary time:
\begin{align} 
M(\mathbf{k},\omega) & = -\frac{1}{\pi} \text{Im} \left[ \mathcal{F}(\mathbf{k},\omega + i\delta) \right], \label{MFromF} \\
\mathcal{F}(\mathbf{k},i \omega_n) & = - \int_0^{\hbar \beta} \left\langle \hat{\mathcal{T}} \hat{\mathcal{F}}_{-\mathbf{k}}(\tau) \hat{\mathcal{F}}_{\mathbf{k}}(0) \right\rangle e^{-i \omega_n \tau} d\tau, \label{FimagDef}
\end{align}
where $\omega_n = 2\pi n/\hbar \beta$ are the bosonic Matsubara frequencies. The time ordering allows one to analyze this Green's function using a diagrammatic expansion. If both anharmonic interactions are neglected, i.e. $\mathcal{V}^{(0)}_{\mathbf{k},\mathbf{q}} = \mathcal{V}^{(1)}_{\mathbf{k},\mathbf{q}} = 0$, the Matsubara Green's function $\mathcal{F}(\mathbf{k},i \omega_n)$ can be calculated exactly by plugging \eqref{FAuxDef} into \eqref{FimagDef}:
\begin{equation}
\mathcal{F}(\mathbf{k},i \omega_n) = \left| \mathcal{V}^{(F)}_{\mathbf{k}} \right|^2 \mathcal{D}_0(\mathbf{k}, i \omega_n),
\end{equation}
where the phonon Green's function $\mathcal{D}_0(\mathbf{k}, i \omega_n)$ is defined as \cite{Mahan2000}:
\begin{equation}
\mathcal{D}_0(\mathbf{k}, i \omega_n) = - \int_0^{\hbar \beta} \left\langle \hat{\mathcal{T}} \left( \opc{-k}(\tau) + \op{k}(\tau) \right) \left( \opc{k}(0) + \op{-k}(0) \right) \right\rangle e^{-i \omega_n \tau} d\tau = \frac{2 \omegaLO}{(i\omega_n)^2 - \omegaLO^2}.
\end{equation}
\begin{figure}
\centering
\includegraphics[width=8.6cm]{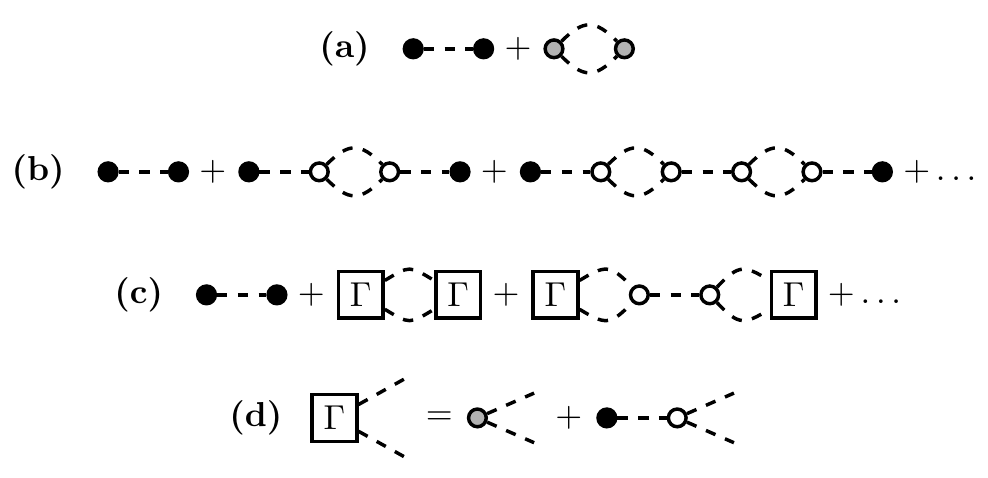}
\caption{\label{fig:DgrmMContributions} Contributions to the Matsubara function $\mathcal{F}(\mathbf{k},i\omega_n)$ considered in this article. Dashed lines represent phonons. Fr\"ohlich interactions $\mathcal{V}_{\mathbf{k}}^{(F)}$ are represented as solid vertices, 1-electron-2-phonon interactions $\mathcal{V}^{(1)}_{\mathbf{k},\mathbf{q}}$ as shaded vertices, and 3-phonon interactions $\mathcal{V}^{(0)}_{\mathbf{k},\mathbf{q}}$ as open vertices. \textbf{(a)} Contributions when the 3-phonon interactions are neglected: only two diagrams are possible, and the result is exact. \textbf{(b)} Contributions when the 1-electron-2-phonon interactions are neglected, which form a Dyson series. \textbf{(c)} Contributions when neither interaction is neglected, expressed in terms of the combined vertex factor $\Gamma$ defined in \textbf{(d)} and equation \eqref{GammaDef}.}
\end{figure}
In the case where the anharmonic terms in the Hamiltonian are not neglected, $\mathcal{F}(\mathbf{k},i \omega_n)$ can be calculated using a diagrammatic expansion. The diagrams that will be considered in this article are shown in \figref{fig:DgrmMContributions}: here, we will motivate these diagrams by investigating the limits where either the 3-phonon interaction or the 1-electron-2-phonon interaction is negligible. First, let us neglect the 3-phonon interaction in the phonon Hamiltonian \eqref{HamPh}, so that it is harmonic. The 1-electron-2-phonon interaction is included by using expression \eqref{FAuxDef} for $\hat{\mathcal{F}}_{\mathbf{k}}$. Calculating the expectation value in \eqref{FimagDef} using Wick's theorem eventually leads to two contributions:
\begin{equation} \label{F1e2ph}
\mathcal{F}(\mathbf{k},i\omega_n) = \left| \mathcal{V}^{(F)}_{\mathbf{k}} \right|^2 \mathcal{D}_0(\mathbf{k},i\omega_n) + \frac{1}{2} \sum_{\mathbf{q}}\left|\mathcal{V}_{\mathbf{q}-\mathbf{k},\mathbf{q}}^{(1)} \right|^2 \left(\frac{-1}{\hbar \beta}  \sum_{m} \mathcal{D}_0(\mathbf{k}-\mathbf{q},i\omega_n-i\omega_m)\mathcal{D}_0(\mathbf{q},i\omega_m) \right).
\end{equation}
This result can be written as the sum of two diagrams, which are shown in \figref{fig:DgrmMContributions}a. The above result is exact when $\mathcal{T}_0 = 0$: no Dyson summation is necessary.

Next, let us only consider the 3-phonon terms as in \figref{fig:DgrmMContributions}b and neglect the 1-electron-2-phonon term in \eqref{FAuxDef}. Then, equation \eqref{FimagDef} for $\mathcal{F}(\mathbf{k},i \omega_n)$ reduces to the full phonon Green's function:
\begin{equation} \label{F3ph}
\mathcal{F}^{(\text{3ph})}(\mathbf{k},i\omega_n) = \left| \mathcal{V}^{(F)}_{\mathbf{k}} \right|^2 \frac{ \mathcal{D}_0(\mathbf{k},i\omega_n)}{1-\mathcal{D}_0(\mathbf{k},i\omega_n)\Pi(\mathbf{k},i\omega_n)} .
\end{equation}
where $\Pi(\mathbf{k},i\omega_n)$ is the self energy of the phonon propagator. It is difficult to calculate in general. In order to proceed analytically, we will approximate it to lowest order: $\Pi(\mathbf{k},i\omega_n) \approx \Pi_0(\mathbf{k},i\omega_n) + O(\mathcal{T}_0^4)$, where $\Pi_0(\mathbf{k},i\omega_n)$ is a bubble diagram. Using the Feynman rules and vertex factors in \cite{Houtput2021}, or by doing the Wick expansion explicitly, this bubble diagram can be evaluated exactly:
\begin{align}
 \Pi_0(\mathbf{k},i\omega_n) & = \vcenter{\hbox{\includegraphics[scale=1]{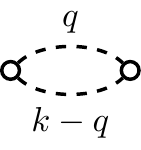}}}, \\
 & = \frac{\omegaLO^2}{2}  \sum_{\mathbf{q}}\left|\frac{\mathcal{V}_{\mathbf{k},\mathbf{q}}^{(0)}}{\hbar \omegaLO} \right|^2 \left(\frac{-1}{\hbar \beta}  \sum_{m} \mathcal{D}_0(\mathbf{k}-\mathbf{q},i\omega_n-i\omega_m)\mathcal{D}_0(\mathbf{q},i\omega_m) \right) \label{Pi0Integrals}.
\end{align}
The Matsubara summation over $m$ can be evaluated by complex integration using a contour that encircles the whole complex plane:
\begin{equation}
\frac{-1}{\hbar \beta}  \sum_{m} \mathcal{D}_0(\mathbf{k}-\mathbf{q},i\omega_n-i\omega_m)\mathcal{D}_0(\mathbf{q},i\omega_m) = \coth\left( \frac{\hbar \beta \omegaLO}{2} \right) \frac{4 \omegaLO}{(i \omega_n)^2 - 4 \omegaLO^2}.
\end{equation}
This result does not depend on $\mathbf{q}$. The remaining sum over $\mathbf{q}$ in \eqref{Pi0Integrals} has been evaluated in \cite{Houtput2021}:
\begin{align}
\sum_{\mathbf{q}}\left|\frac{\mathcal{V}_{\mathbf{k},\mathbf{q}}^{(0)}}{\hbar \omegaLO} \right|^2 = \frac{4 \mathcal{T}_0^2}{15 \tilde{V}_0},
\end{align}
where $\tilde{V}_0 = V_0 \left(\frac{2m_b \omegaLO}{\hbar}\right)^{\frac{3}{2}}$ is a dimensionless parameter representing the size of the unit cell. Therefore, the lowest order self energy of the phonon propagator is equal to:
\begin{equation} \label{Pi0Solution}
\Pi_0(\mathbf{k},i\omega_n) = \omegaLO^2 \frac{2 \mathcal{T}_0^2}{15 \tilde{V}_0} \coth\left(\frac{\hbar \beta \omegaLO}{2} \right) \frac{4 \omegaLO}{(i\omega_n)^2 - 4 \omegaLO^2}.
\end{equation}
With this expression for the self energy, equation \eqref{F3ph} corresponds to the Dyson series in \figref{fig:DgrmMContributions}. 

In order to include both the 3-phonon interactions and the 1-electron-2-phonon interactions, we consider the same Dyson series as in \figref{fig:DgrmMContributions}a, but starting and ending with two possible ways for the electron to create two phonons, as in \figref{fig:DgrmMContributions}b. Firstly, the electron can create a phonon through the Fr\"ohlich interaction, which then splits into two phonons through the 3-phonon interaction. Secondly, the electron can simultaneously create two phonons through the 1-electron-2-phonon interaction. These two processes are due to the first and second term in \eqref{FAuxDef}, respectively; in the derivation of \eqref{F3ph}, only the first term was considered. Both processes can be combined into a single vertex $\Gamma$, as in figure \figref{fig:DgrmMContributions}c:
\begin{align}
\Gamma(\mathbf{k}-\mathbf{q},i \nu_n - i \nu_{m};\mathbf{q}, i \nu_{m}) & = \mathcal{V}^{(1)}_{\mathbf{q},\mathbf{q}-\mathbf{k}} + \frac{1}{\hbar} \mathcal{V}^{(F)}_{\mathbf{k}} \mathcal{V}^{(0)}_{\mathbf{k},\mathbf{q}} \mathcal{D}_0(\mathbf{k},i \nu_n), \label{GammaDef} \\
& = -\frac{\mathcal{V}^{(F)}_{\mathbf{k}} \mathcal{V}^{(0)}_{\mathbf{k},\mathbf{q}}}{\hbar \omegaLO} \left( \frac{\mathcal{T}_1}{\mathcal{T}_0} - \omegaLO \mathcal{D}_0(\mathbf{k}, i \nu_n) \right).
\end{align}
Summing all the terms in \figref{fig:DgrmMContributions}c gives the final expression for the Matsubara Green's function:
\begin{equation} \label{Ffinal}
\mathcal{F}(\mathbf{k},i\omega_n) = \left| \mathcal{V}^{(F)}_{\mathbf{k}} \right|^2 \left( \mathcal{D}_0(\mathbf{k},i\omega_n) + \frac{\left(\frac{\mathcal{T}_1}{\omegaLO \mathcal{T}_0} -  \mathcal{D}_0(\mathbf{k},i\omega_n)\right)^2 \Pi_0(\mathbf{k},i\omega_n)}{1-\mathcal{D}_0(\mathbf{k},i\omega_n)\Pi_0(\mathbf{k},i\omega_n)} \right).
\end{equation}
This is a closed expression for the Matsubara Green's function since the lowest order phonon self energy $\Pi_0(\mathbf{k},i\omega_n)$ is known \eqref{Pi0Solution}. Equation \eqref{Ffinal} contains both \eqref{F1e2ph} and \eqref{F3ph} as limiting cases: this can be seen by taking the limit $\mathcal{T}_0 \rightarrow 0$ or $\mathcal{T}_1 \rightarrow 0$. Graphically, this procedure corresponds to starting from \ref{fig:DgrmMContributions}c, and removing either the 3-phonon vertex or the 1-electron-2-phonon vertex: this indeed results in \ref{fig:DgrmMContributions}a or \ref{fig:DgrmMContributions}b, respectively. Expression \eqref{Ffinal} is valid up to second order in $\mathcal{T}_0$. To get a more accurate result the full phonon self energy $\Pi(\mathbf{k},i\omega_n)$ should be used in \eqref{Ffinal}, but in order to continue the analytic treatment of this article we will be satisfied with the results up to second order in $\mathcal{T}_0$.

To calculate the phonon spectral function $M(\mathbf{k},\omega)$, we use expression \eqref{Pi0Solution} for the phonon self energy in \eqref{Ffinal}, and split the resulting Matsubara Green's function into partial fractions. This yields:
\begin{equation} \label{FSplit}
\mathcal{F}(\mathbf{k},i\omega_n) = \left| \mathcal{V}^{(F)}_{\mathbf{k}} \right|^2 \left[c_1 \frac{2 \omegaLO x_1}{(i\omega_n)^2 - \omegaLO^2 x_1^2} + c_2 \frac{2 \omegaLO x_2}{(i\omega_n)^2 - \omegaLO^2 x_2^2} \right],
\end{equation}
where the numerical constants $x_1$, $x_2$, $c_1$ and $c_2$ are defined as the following combinations of $\mathcal{T}_0$, $\mathcal{T}_1$, $\tilde{V}_0$, and $\hbar \beta \omegaLO$:
\begin{align}
x_1 & := \sqrt{\frac{1}{2} \left(5 - 3 \sqrt{1+\frac{64 \mathcal{T}_0^2}{135 \tilde{V}_0} \coth\left(\frac{\hbar \beta \omegaLO}{2} \right)} \right)}, \label{x1FiniteTemp} \\
x_2 & := \sqrt{\frac{1}{2} \left(5 + 3 \sqrt{1+\frac{64 \mathcal{T}_0^2}{135 \tilde{V}_0} \coth\left(\frac{\hbar \beta \omegaLO}{2} \right)} \right)}, \label{x2FiniteTemp}
\end{align}
\begin{align}
c_1 & := \frac{1}{2 x_1} \left( \begin{array}{l}
1 + \frac{4 \mathcal{T}_1^2}{15 \tilde{V}_0} \coth\left(\frac{\hbar \beta \omegaLO}{2} \right) \\
 + \left[1 + \left(\frac{8}{3} \mathcal{T}_0 \mathcal{T}_1 - \mathcal{T}_1^2 \right)\frac{4}{15 \tilde{V}_0} \coth\left(\frac{\hbar \beta \omegaLO}{2} \right) \right] \frac{1}{\sqrt{1+\frac{64 \mathcal{T}_0^2}{135 \tilde{V}_0} \coth\left(\frac{\hbar \beta \omegaLO}{2} \right)}}
\end{array}  \right), \label{c1FiniteTemp} \\
c_2 & := \frac{1}{2 x_2} \left( \begin{array}{l}
1 + \frac{4 \mathcal{T}_1^2}{15 \tilde{V}_0} \coth\left(\frac{\hbar \beta \omegaLO}{2} \right) \\
 - \left[1 + \left(\frac{8}{3} \mathcal{T}_0 \mathcal{T}_1 - \mathcal{T}_1^2 \right)\frac{4}{15 \tilde{V}_0} \coth\left(\frac{\hbar \beta \omegaLO}{2} \right) \right] \frac{1}{\sqrt{1+\frac{64 \mathcal{T}_0^2}{135 \tilde{V}_0} \coth\left(\frac{\hbar \beta \omegaLO}{2} \right)}}
\end{array}  \right). \label{c2FiniteTemp}
\end{align}
The phonon spectral function $M(\mathbf{k},\omega)$ can then be straightforwardly calculated from \eqref{MFromF}. Since:
\begin{equation}
\lim_{\epsilon \rightarrow 0^+} -\frac{1}{\pi} \text{Im}\left[\frac{2\nu}{(\omega + i \epsilon)^2 - \nu^2} \right] = \delta(\omega - \nu) - \delta(\omega + \nu),
\end{equation}
for any frequency $\nu$, the phonon spectral function becomes:
\begin{equation}
M(\mathbf{k},\omega) = \left|\mathcal{V}^{(F)}_{\mathbf{k}}\right|^2 \left[ \begin{array}{l}
c_1 \left(\delta(\omega-\omegaLO x_1) - \delta(\omega+\omegaLO x_1)\right) \\
 + c_2 \left(\delta(\omega-\omegaLO x_2) - \delta(\omega+\omegaLO x_2)\right)
\end{array} \right] + O(\mathcal{T}_1^4). \label{MGeneral}
\end{equation}
The phonon spectral function in the region $\omega > 0$ is therefore a sum of two infinitely sharp delta peaks. In the absence of 3-phonon interaction, these peaks appear at $\omega = \omegaLO$ and $\omega = 2\omegaLO$, and can be associated with the Fr\"ohlich interaction and the 1-electron-2-phonon interaction respectively. The 3-phonon terms only shift the locations and heights of these delta peaks. Indeed, the delta peaks occur at $\omega = x_1 \omegaLO$ and $\omega = x_2 \omegaLO$, or equivalently:
\begin{align}
\omega & \approx \omegaLO -  \mathcal{T}_0^2 \frac{8 \omegaLO}{45 \tilde{V}_0} \coth\left(\frac{\hbar \beta \omegaLO}{2} \right) + O(\mathcal{T}_0^4), \\
\text{and }\omega & \approx 2\omegaLO + \mathcal{T}_0^2 \frac{4 \omegaLO}{45 \tilde{V}_0} \coth\left(\frac{\hbar \beta \omegaLO}{2} \right) + O(\mathcal{T}_0^4).
\end{align}
Other treatments of 3-phonon anharmonicity \cite{Akhieser1939, Carruthers1962, Klemens1966, Ushioda1972, Lockwood2005, Setty2020} usually lead to a finite lifetime of the phonon, which dampens and broadens the delta peaks in the spectral function $M(\mathbf{k},\omega)$. This does not happen in \eqref{MGeneral} because the Hamiltonian \eqref{HamTot}-\eqref{HamElPh} of \cite{Houtput2021} only includes one longitudinal optical phonon mode, and neglects all other modes. The only possible 3-phonon processes LO $\rightarrow$ LO + LO and LO + LO $\rightarrow$ LO do not satisfy conservation of energy, since the initial (resp. final) state has an energy of $E_i = \hbar \omegaLO$ whereas the final (resp. initial) state has energy $E_f = 2 \hbar \omegaLO$. Therefore, according to Fermi's golden rule \cite{LandauLifshitz3}:
 \begin{equation}
\frac{1}{\tau} \sim \frac{2\pi}{\hbar} \sum_f \left|\brakett{f}{\hat{H}_{3-\text{ph}}}{i}\right|^2 \delta(E_f - E_i),
\end{equation} 
this process cannot contribute to the finite lifetime of the phonon \cite{Srivastava2019}. Other 3-phonon processes, such as LO $\rightarrow$ LA + LA, would indeed lead to a finite lifetime \cite{Klemens1966}. The fact that no broadening is present is a significant limitation of the 3-phonon interaction term \eqref{HamPh}. Therefore, for the discussion of the results in \secref{sec:Results}, we will mainly focus on the effect of the 1-electron-2-phonon interaction, rather than the effect of the LO $\rightarrow$ LO + LO 3-phonon process.

Because the phonon spectral function is composed of delta peaks, the integral in expression \eqref{SigmaImGen} for the approximate memory function $\text{Im}\left[\Sigma_0(\omega)\right]$ can be calculated explicitly. Since $M(\mathbf{k},\omega) \sim \left| \mathcal{V}^{(F)}_{\mathbf{k}} \right|^2$, and the dynamical structure factor $S(\mathbf{k},\omega) = S(k, \omega)$ is isotropic in $\mathbf{k}$ for the homogeneous electron gas, the sum over $\mathbf{k}$ in this expression will always be of the following form:
\begin{equation}
\sum_{\mathbf{k}} k^2 \left| \mathcal{V}^{(F)}_{\mathbf{k}} \right|^2 S(\mathbf{k},\omega) = \frac{\alpha}{2 \pi^2} (\hbar\omegaLO)^2 \sqrt{\frac{\hbar}{2 m_b \omegaLO}} \times 4\pi \int S(k,\omega) k^2 \tildeff k.
\end{equation}
Therefore, for all further results, we require the integral of the dynamical structure factor over all momenta. A straightforward calculation leads to the following expression for the imaginary part of $\Sigma_0(\omega)$:
\begin{small}
\begin{equation} \label{ImSigma0Useful}
\text{Im}\left[\Sigma_0(\omega)\right] = - \frac{4 \alpha}{3} \left( \frac{\hbar \omegaLO}{2m_b} \right)^{\frac{3}{2}} \sum_{i=1}^2 \sum_{\pm} \frac{c_i}{\omega} [1 + n_B(\pm x_i \omegaLO) + n_B(\omega \mp x_i \omegaLO) ] \int S(k, \omega \mp x_i \omegaLO) k^2\tildeff k,
\end{equation}
\end{small}where $x_i$ and $c_i$ are given by expressions \eqref{x1FiniteTemp}-\eqref{c2FiniteTemp}.

The calculation of the structure factor will be discussed in the next section. Once the integrated structure factor $\int S(k, \omega) k^2\tildeff k$ has been calculated, the conductivity $\sigma(\omega)$ can be found by using equation \eqref{ImSigma0Useful} to find the imaginary part of $\Sigma_0(\omega)$, equation \eqref{SigmaReGen} to find its real part, equation \eqref{MemoryFullDef} to find the memory function, and equation \eqref{condMemory} to find the conductivity.

Note that \eqref{ImSigma0Useful} reproduces the result for the Fr\"ohlich polaron gas in \cite{Tempere2001} if we set $\mathcal{T}_0 = \mathcal{T}_1 = 0$ and work in the zero temperature limit $\beta \rightarrow +\infty$:
\begin{equation} \label{imsigmaT0}
\text{Im}[\Sigma_0(\omega)] = -\frac{\pi}{3 m_b \hbar \omega} \sum_{\mathbf{k}}  k^2 \left| \mathcal{V}^{(F)}_{\mathbf{k}} \right|^2 S(\mathbf{k},\omega - \omegaLO) \Theta(\omega - \omegaLO).
\end{equation}
where $\Theta(x)$ is the Heaviside function. Combining equations \eqref{condMemory}, \eqref{MemoryFullDef}, \eqref{SigmaReGen}, and the fact that $\Sigma_0(0) = 0$ at zero temperature according to equations \eqref{SigmaReGen} and \eqref{imsigmaT0}, eventually yields:
\begin{equation}
\text{Re}[\sigma(\omega)] = \frac{\pi n e^2}{3 m_b^2 \hbar \omega^3} \sum_{\mathbf{k}}  k^2 \left| \mathcal{V}^{(F)}_{\mathbf{k}} \right|^2 S(\mathbf{k},\omega - \omegaLO) \Theta(\omega - \omegaLO).
\end{equation}
which is indeed the result of \cite{Tempere2001}, up to a conventional factor $\pi$ that is included in the definition of $S(\mathbf{k},\omega)$ in \cite{Tempere2001}. The method presented in this chapter is therefore an extension of the method in \cite{Tempere2001}: the treatment of this chapter includes finite temperatures as well as 1-electron-2-phonon interaction and 3-phonon interactions of the form \eqref{HamPh}-\eqref{HamElPh}.

\subsection{Dynamical structure factor of the electron gas} \label{sec:DynamicalStructureFactor}
The dynamical structure factor $S(\mathbf{k},\omega)$ of the homogeneous electron gas, defined by \eqref{StructureFactorDef}, is a well-known quantity in the literature \cite{Devreese1980_1, Devreese1980_2, Mahan2000, Ancarani2016}. It represents the response of the homogeneous electron gas to a perturbation with momentum $\hbar \mathbf{k}$ and energy $\hbar \omega$, and is related to its dielectric function $\varepsilon(\mathbf{k},\omega)$ as follows \cite{Mahan2000}:
\begin{equation}
S(\mathbf{k},\omega) = -\frac{\hbar \varepsilon_{\text{vac}} \varepsilon_{\infty} k^2}{\pi n e^2} \text{Im}\left[\frac{1}{\varepsilon(\mathbf{k},\omega)} \right].
\end{equation}
The structure factor depends on the density $n$ of the electron gas. In this article, the dependence on the density is written in terms of the Wigner-Seitz radius \cite{Mahan2000} $r_s = \frac{1}{\varepsilon_{\infty} a_B} \left( \frac{3}{4 \pi n} \right)^{\frac{1}{3}}$ where $a_B = 0.53 \text{\AA}$ is the Bohr radius. The dynamical structure factor is written in terms of standard quantities derived from the density: the Fermi wavevector $k_F = (3 \pi^2 n)^{1/3}$, the Fermi energy $E_F = \frac{\hbar^2 k_F^2}{2m_b}$, and the plasma frequency $\omega_{\text{pl}} = \frac{n e^2}{m_b \varepsilon_{\text{vac}}}$ \cite{Mahan2000}.

There is no exact expression for the dynamical structure factor. However, there are several models that describe the dynamical structure factor with increasing degree of precision. Most of these models can be expressed in terms of the Lindhard polarization function \cite{Mahan2000}:
\begin{equation} \label{LindhardP}
P(\mathbf{k},\omega) = \lim_{\delta \rightarrow 0^+} \mathcal{V}^{(C)}_{\mathbf{k}} \sum_{\mathbf{q}} \frac{n_F(\varepsilon_{\mathbf{q}}) - n_F(\varepsilon_{\mathbf{q}+\mathbf{k}})}{\varepsilon_{\mathbf{q}} - \varepsilon_{\mathbf{q}+\mathbf{k}} + \hbar (\omega + i \delta)} := A(\mathbf{k},\omega) + i B (\mathbf{k},\omega),
\end{equation}
where $\varepsilon_{\mathbf{k}}$ and $\mathcal{V}^{(C)}_{\mathbf{k}}$ are given by \eqref{epsilonK} and \eqref{VCoulomb}, and $n_F(E) = 1/(e^{\beta (E - \mu)}+1)$ is the Fermi-Dirac distribution. The real and imaginary parts $A(\mathbf{k},\omega)$ and $B(\mathbf{k},\omega)$ can be found by evaluating \eqref{LindhardP}, or by using the expressions found in \cite{Mahan2000, Devreese2010}.

In this article, we will discuss and compare the following four commonly used \cite{Mahan2000} models for the dynamical structure factor, listed in order of increasing precision:
\begin{itemize}
\item The one polaron model, which neglects the Coulomb interaction and exchange effects of the electron gas \cite{Tempere2001}:
\begin{equation} \label{SFonepolaron}
S(\mathbf{k},\omega) = \delta\left(\omega - \frac{\hbar k^2}{2 m_b} \right).
\end{equation}
This basic model allows for further analytical calculations due to its simplicity.
\item The Hartree-Fock model, which includes the exchange effects but neglects the Coulomb interaction:
\begin{equation} \label{SFHartreeFock}
S(\mathbf{k},\omega) = -\frac{\hbar \varepsilon_{\text{vac}} \varepsilon_{\infty} k^2}{\pi n e^2} B(\mathbf{k},\omega).
\end{equation}
This model is valid when the Coulomb interaction between the electrons is negligible, which is the case when $\hbar \omega_{\text{pl}} \ll E_F$.
\item The Lindhard model, also known as the Random Phase Approximation or RPA model, which includes the Coulomb interaction up to lowest order \cite{Lindhard1954}:
\begin{equation} \label{SFLindhard}
S(\mathbf{k},\omega) = -\frac{\hbar \varepsilon_{\text{vac}} \varepsilon_{\infty} k^2}{\pi n e^2} \frac{B(\mathbf{k},\omega)}{[1-A(\mathbf{k},\omega)]^2 + B(\mathbf{k},\omega)^2}.
\end{equation}
\item The Hubbard model, which includes a local field factor $G(k)$ to account for the exchange and correlation hole around the electron \cite{Mahan2000}:
\begin{equation}
G(k) = \frac{1}{2} \frac{k^2}{k^2 + k_F^2 + \frac{3 m_b \omega_{\text{pl}}^2}{2 E_F}}.
\end{equation}
The Hubbard dynamical structure factor is then given by:
\begin{equation} \label{SFHubbard}
S(\mathbf{k},\omega) = -\frac{\hbar \varepsilon_{\text{vac}} \varepsilon_{\infty} k^2}{\pi n e^2} \frac{B(\mathbf{k},\omega)}{[1-(1-G(k))A(\mathbf{k},\omega)]^2 + (1-G(k)) B(\mathbf{k},\omega)^2}.
\end{equation}
The Hubbard model is especially good for the large polaron problem, since the local field factor $G(k)$ is small in the $k \rightarrow 0$ limit. Therefore, although more specialized models for the structure factor exist \cite{Devreese1980_1, Devreese1980_2}, we will limit ourselves to the RPA and Hubbard models in this article - this assumption will be motivated \emph{a posteriori} in \secref{sec:Results}.
\end{itemize}
\begin{figure}
\centering
\includegraphics[width=8.6cm]{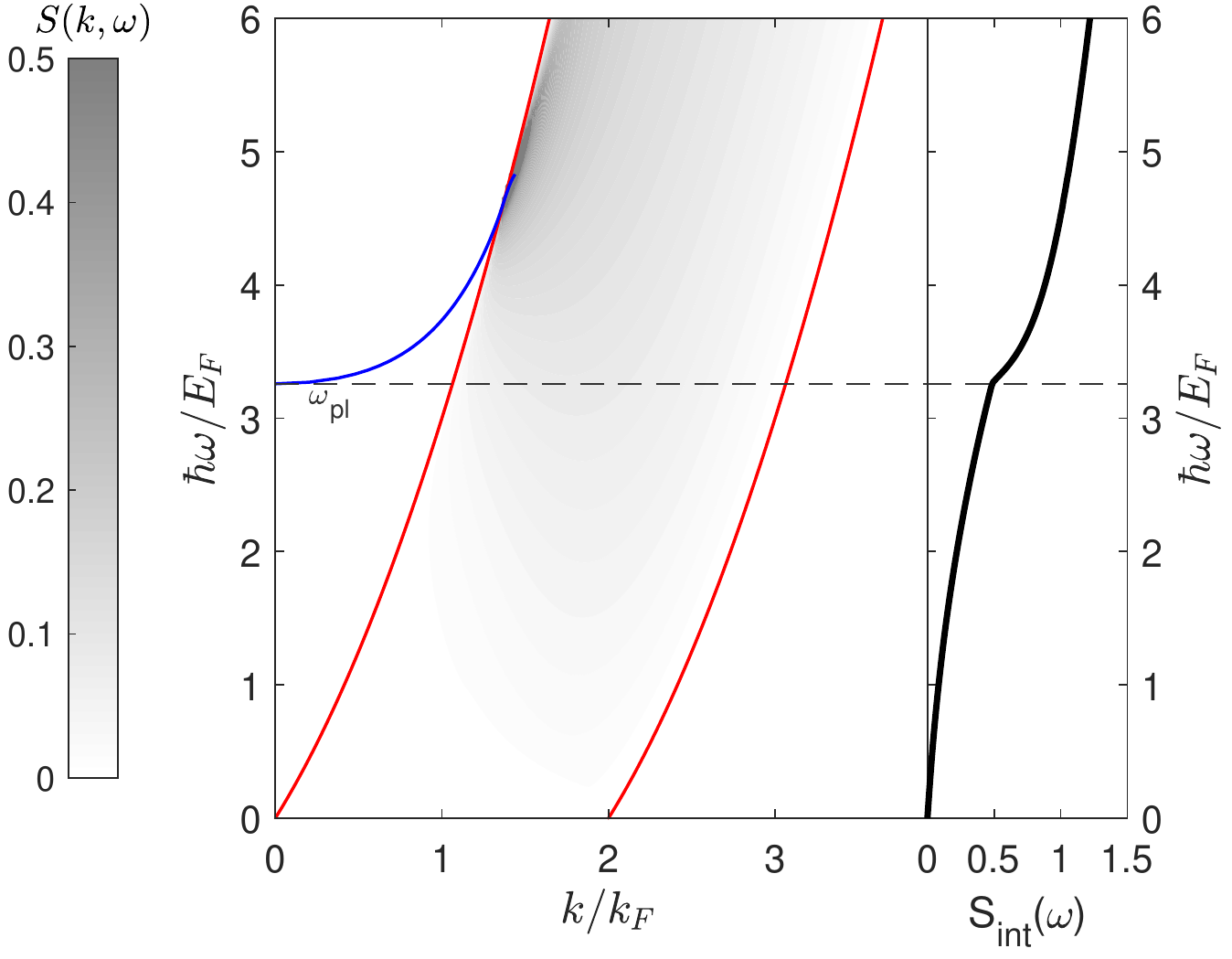}
\caption{\label{fig:StructureHubbard} Contour plot of the dynamical structure factor $S(\mathbf{k},\omega)$ of the electron gas, calculated at zero temperature using the Hubbard model \eqref{SFHubbard} with $r_s = 12$. Outside the red lines, the structure factor is  exactly zero, except for the Dirac delta peak due to the formation of undamped plasmons (blue). The right inset shows the structure factor integrated over all momenta, defined as $S_{\text{int}}(\omega) = \frac{E_F}{\hbar k_F^3}\int S(k,\omega)k^2 \tildeff k$.}
\end{figure}
\figref{fig:StructureHubbard} shows the dynamical structure factor for the Hubbard model. To calculate the conductivity of the polaron gas \eqref{ImSigma0Useful}, the structure factor must be integrated over all momenta $k^2 \tildeff k$. The dynamical structure factor has a sharp undamped plasmon peak \cite{Tempere2001} when $\varepsilon(\mathbf{k},\omega) = 0$, which needs to be treated carefully when performing this integral. The plasmons lead to a kink at $\omega_{\text{pl}}$ in the integral of the structure factor, which will lead to additional features in the optical conductivity \cite{Tempere2001}.

\section{Results} \label{sec:Results}

\subsection{Material parameters} \label{sec:MaterialParameters}
\begin{table}
\centering
\begin{tabular}{r|cccc|ccccc}
 & $\omegaLO$ (THz) & $m_b$ ($m_e$) & $\varepsilon_0$ & $\varepsilon_{\infty}$ & $\alpha$ & $\mathcal{T}_1$ & $\tilde{V}_0$ & $\frac{E_{\text{Ry}}}{\hbar\omegaLO}$ \\ \hline
\textbf{BN} & 38.41 & 0.329 & 6.98 & 4.62 & 0.973 & -0.00134 & 0.00121 & 8.28 \\
\textbf{BP} & 24.48 & 0.331 & 9.28 & 9.19 & 0.018 & -0.00085 & 0.00123 & 3.44 \\
\textbf{AlN} & 26.52 & 0.285 & 8.59 & 4.62 & 1.492 & -0.00069 & 0.00100 & 10.42 \\
\textbf{AlP} & 14.65 & 0.311 & 10.41 & 8.14 & 0.561 & 0.00050 & 0.00092 & 6.62 \\
\end{tabular}
\caption{\label{tab:MaterialParameters} Material parameters for electron polarons in several cubic III-V semiconductors The material parameters in the left part of the table are found in \cite{MaterialsProject_BN, MaterialsProject_BP, MaterialsProject_AlN, MaterialsProject_AlP, Ricci2017}; the other parameters are derived quantities, except for $\mathcal{T}_1$ which were calculated by \cite{ViennaCommunication}.}
\end{table}

The combination of expressions \eqref{condMemory}, \eqref{MemoryFullDef}, \eqref{SigmaReGen}, \eqref{ImSigma0Useful}, and any of the structure factors \eqref{SFonepolaron}-\eqref{SFHubbard} allows us to calculate the conductivity of the anharmonic polaron gas, given values for all the necessary material parameters. At the very least, this includes values of the electron density $n$ or the Wigner-Seitz radius $r_s$, and the polaron material parameters $\alpha$, $\mathcal{T}_0$, $\mathcal{T}_1$, and $\tilde{V}_0$. The phonons and electron gas have their own characteristic energy scales $\hbar \omegaLO$ and $E_F$, so the result will also depend on the ratio of these energies. This parameter can be written in terms of the Rydberg energy $E_{\text{Ry}}$:
\begin{align}
\frac{E_F}{\hbar \omegaLO} & = \frac{E_{\text{Ry}}}{\hbar \omegaLO} \left( \frac{9\pi}{4} \right)^{\frac{2}{3}} \frac{1}{r_s^2}, \\
\frac{E_{\text{Ry}}}{\hbar \omegaLO} & = \left( \frac{\alpha}{1 - \frac{\varepsilon_{\infty}}{\varepsilon_0}} \right)^2.
\end{align}
$E_{\text{Ry}}/\hbar \omegaLO$ is independent of the electron density, and is therefore another dimensionless material parameter. Finally, the result also depends on the temperature of the system.
\tabref{tab:MaterialParameters} shows the values of $\alpha$, $\mathcal{T}_1$, $\tilde{V}_0$, and $\frac{E_{\text{Ry}}}{\hbar\omegaLO}$ for the lightest III-V semiconductors, for which the Hamiltonian \eqref{HamTot}-\eqref{HamElPh} is valid. The values of $\mathcal{T}_1$ for these four materials were calculated by \emph{ab initio} methods \cite{ViennaCommunication}. These materials turn out to have negligible 1-electron-2-phonon interaction since their values of $\mathcal{T}_1$ are quite low $(\mathcal{T}_1 \sim 10^{-3})$. Currently, the values of $\mathcal{T}_1$ are unknown for all other materials. In the remainder of this section, we will therefore use larger, arbitrarily chosen values for $\mathcal{T}_1$ to demonstrate the effect of significant 1-electron-2-phonon interaction. 
Similarly, $\mathcal{T}_0$ is not known for any material to the best of our knowledge. Comparisons with other treatments of 3-phonon anharmonicity \cite{Akhieser1939, Carruthers1962, Klemens1966, Ushioda1972, Lockwood2005, Setty2020, Srivastava2019} are difficult for the reasons outlined in \secref{sec:PhononSpectralFunction}: $\mathcal{T}_0$ represents the strength of only the LO $\rightarrow$ LO + LO process, whereas most treatments take all possible 3-phonon processes into account at once. Therefore, we will choose arbitrary values of $\mathcal{T}_0$ to show the effect of the 3-phonon interaction.

The Wigner-Seitz radius $r_s$ is fully determined by the carrier density $n$, which can be chosen freely in experiments by doping. For concreteness, results in this article will be plotted using $n \sim 5 \times 10^{18} \text{cm}^{-3}$, a typical density for doped semiconductors \cite{Tempere2001}. Using the values in \tabref{tab:MaterialParameters}, this corresponds to a Wigner-Seitz unit radius of the order $r_s \sim 12$.

\subsection{The low-frequency and high-frequency limits, and the imaginary part of the conductivity}
Using the theory of \secref{sec:Theory}, the optical conductivity of the anharmonic polaron gas can be calculated in several different limits. Here, the low-frequency and high-frequency limits are explored, and we show that both limits can be written as a Drude conductivity \cite{Ashcroft1976}. Combining \eqref{condMemory} and \eqref{MemoryFullDef} allows us to write the optical conductivity $\sigma(\omega)$ in terms of $\Sigma_0(\omega)$:
\begin{equation} \label{condSigma0}
\sigma(\omega) = i \frac{n e^2}{m_b} \frac{1}{\omega + \frac{i}{\tau_{\text{eff}}}} \left(1 + \frac{\Sigma_0(\omega) - \Sigma_0(0)}{\omega} \right),
\end{equation}
where $\tau_{\text{eff}}^{-1} = - \text{Im}[\Sigma_0(0)]$ plays the role of collision rate of the electrons and the phonons. From expressions \eqref{SigmaReGen} and \eqref{SigmaImGen}, it can be seen that the imaginary part of $\Sigma_0(\omega)$ is an even function and its real part is an odd function. Taking the $\omega \rightarrow +\infty$ limit of equation \eqref{condSigma0} yields:
\begin{equation} \label{condDrudeHigh}
\lim_{\omega \rightarrow +\infty} \sigma(\omega) = i \frac{n e^2}{m_b \omega}.
\end{equation}
At high frequencies, the conductivity simply reduces to the Drude conductivity of the free electron gas, as the heavy ions are too slow to follow the fast-moving electrons. Similarly, taking the $\omega \rightarrow 0$ limit of equation \eqref{condSigma0} also yields a Drude conductivity, but with a different effective mass $m_{\text{eff}}$ and with the relaxation time $\tau_{\text{eff}}$:
\begin{equation} \label{condDrude}
\lim_{\omega \rightarrow 0} \sigma(\omega) = i \frac{n e^2}{m_{\text{eff}} \left( \omega + \frac{i}{\tau_{\text{eff}}} \right)}.
\end{equation}
The relaxation time $\tau_{\text{eff}}$ and effective mass $m_{\text{eff}}$ of the conductivity are defined from the low-frequency behavior of the memory function $\Sigma_0$:
\begin{align}
\frac{1}{\tau_{\text{eff}}} & = - \text{Im}[\Sigma_0(0)] = \frac{4 \alpha}{3} \left( \frac{\hbar \omegaLO}{2m_b} \right)^{\frac{3}{2}} \sum_{i=1}^2 \sum_{\pm} c_i n_B'(\pm x_i \omegaLO) \int S(k, \pm x_i \omegaLO) k^2\tildeff k, \label{tauEffDef} \\
\frac{m_b}{m_{\text{eff}}} & = 1 + \text{Re}[\Sigma_0'(0)] = 1 + \frac{2}{\pi} \int_0^{\infty} \frac{\text{Im}[\Sigma_0(\omega)] - \text{Im}[\Sigma_0(0)]}{\omega^2} \tildeff \omega. \label{mEffDef}
\end{align}
For a single polaron, $\tau_{\text{eff}}$ represents the average collision time between collisions with a phonon, and $m_{\text{eff}}$ represents the effective polaron mass. Note that at temperature zero, $\tau^{-1}_{\text{eff}} = 0$ because no phonons are present. Indeed, in expression \eqref{tauEffDef}, the derivative of the Bose-Einstein distribution becomes $n_B'(\pm x_i \omegaLO) \rightarrow - \delta(\pm x_i \omegaLO) = 0$. In practice, this means that the optical conductivity \eqref{condDrude} will have a Dirac delta contribution at temperature zero:
\begin{equation}
\lim_{\omega \rightarrow 0} \sigma(\omega) \big|_{T=0} = \frac{\pi n e^2}{m_{\text{eff}}} \delta(\omega) + i \frac{n e^2}{m_{\text{eff}}  \omega}.
\end{equation}
This contribution is necessary to satisfy the f-sum rule at zero temperature  and is well-known in the literature \cite{Devreese1977, Tempere2001, Alexandrov2010}.

\begin{figure}
\centering
\includegraphics[width=8.6cm]{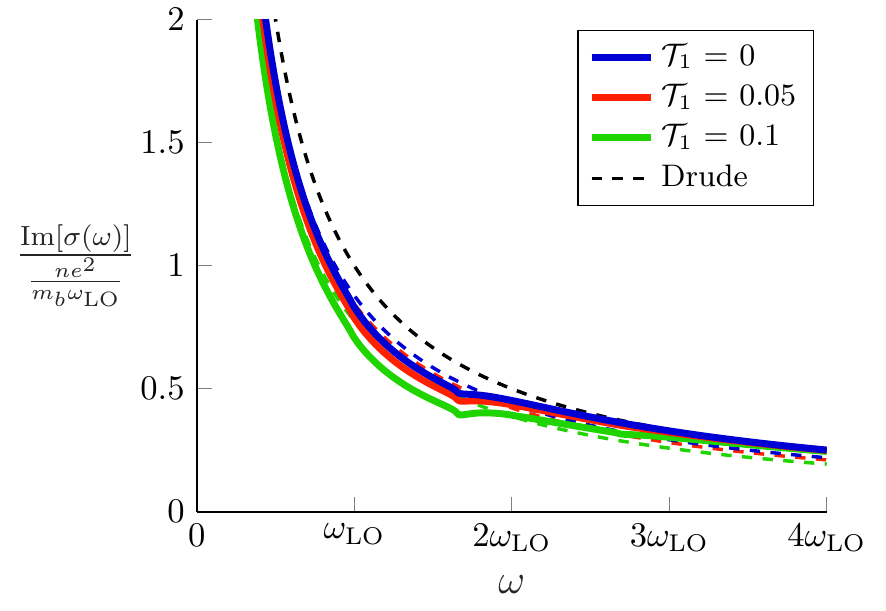}
\caption{\label{fig:CondImag} The imaginary part of the optical conductivity of the polaron gas at zero temperature, calculated using the Hubbard model \eqref{SFHubbard} for the structure factor, $\alpha = 1$, $\tilde{V}_0 = 0.001$, $\mathcal{T}_0 = 0$, $E_{\text{Ry}}/\hbar \omegaLO = 8$, and $r_s = 12$. The dashed lines represent the Drude conductivity: the black dashed line represents the high-frequency limit \eqref{condDrudeHigh} calculated with the band mass, and the colored dashed lines represent the low-frequency limit \eqref{condDrude} calculated with the effective polaron mass.}
\end{figure}
Overall, the imaginary part of the conductivity can be excellently described by a combination of the two limits \eqref{condDrudeHigh}-\eqref{condDrude}, as shown on \figref{fig:CondImag}. There are some features in the intermediate region $\omega \sim \omegaLO$, which becomes more pronounced as $\alpha$ is larger; however, when $\alpha$ is too large, the theory presented in this article becomes invalid and more specialized techniques are necessary \cite{Mishchenko2003}. In the weak-coupling limit $\alpha \rightarrow 0$, the main effect on the imaginary part of the conductivity is to change the effective polaron mass in the low-frequency limit \eqref{condDrude}.

At zero temperature and assuming the single polaron structure factor \eqref{SFonepolaron}, the effective polaron mass can be calculated explicitly using expression \eqref{mEffDef} in combination with \eqref{ImSigma0Useful}. With $n_B(\omega) = \Theta(\omega)-1$ at temperature zero, the resulting inverse effective mass is:
\begin{equation}
\frac{m_b}{m_{\text{eff}}} = 1 - \frac{\alpha}{6} \left( \frac{c_1}{x_1^{3/2}} + \frac{c_2}{x_2^{3/2}} \right) + O(\alpha^2, \mathcal{T}_0^4).
\end{equation}
This result is valid up to second order in $\mathcal{T}_0$ because the phonon self energy \eqref{Pi0Solution} was approximated up to lowest order. Expanding the expressions \eqref{x1FiniteTemp}-\eqref{c2FiniteTemp} for the coefficients $x_1, x_2, c_1, c_2$ up to second order in $\mathcal{T}_0$ yields:
\begin{equation}
\frac{m}{m_{\text{eff}}} = 1-\frac{\alpha}{6} - \frac{1}{90\sqrt{2}} \frac{\alpha}{\tilde{V}_0} \left[ \mathcal{T}_1^2 + \frac{4(4\sqrt{2}-1)}{3} \mathcal{T}_0 \mathcal{T}_1 + \frac{4 (11\sqrt{2}+1)}{9} \mathcal{T}_0^2 \right] + O(\alpha^2, \mathcal{T}_0^4),
\end{equation}
which is the same as the polaron effective mass calculated from perturbation theory \cite{Houtput2021}.

\subsection{Optical absorption of the polaron gas}
\begin{figure}
\includegraphics[width=8.1cm]{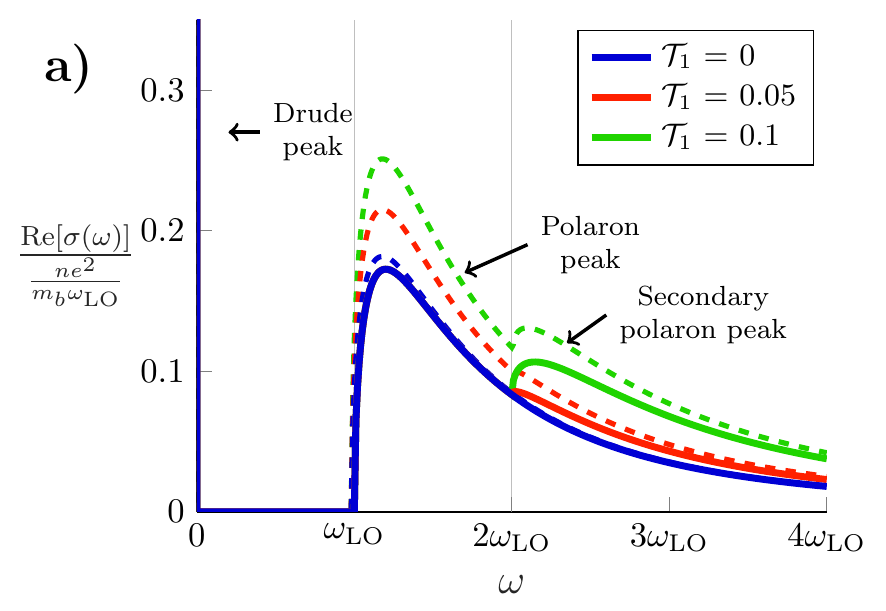}
\includegraphics[width=8.1cm]{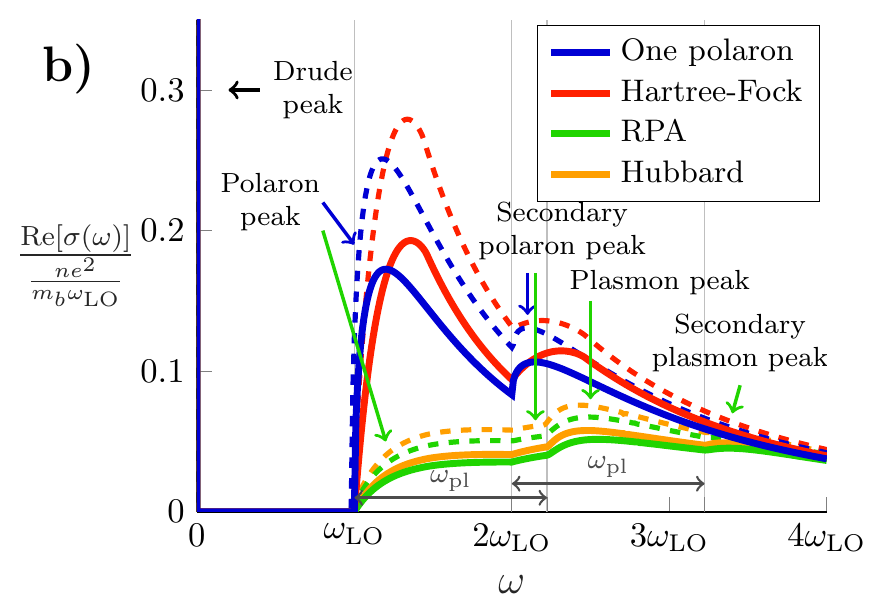}
\includegraphics[width=8.1cm]{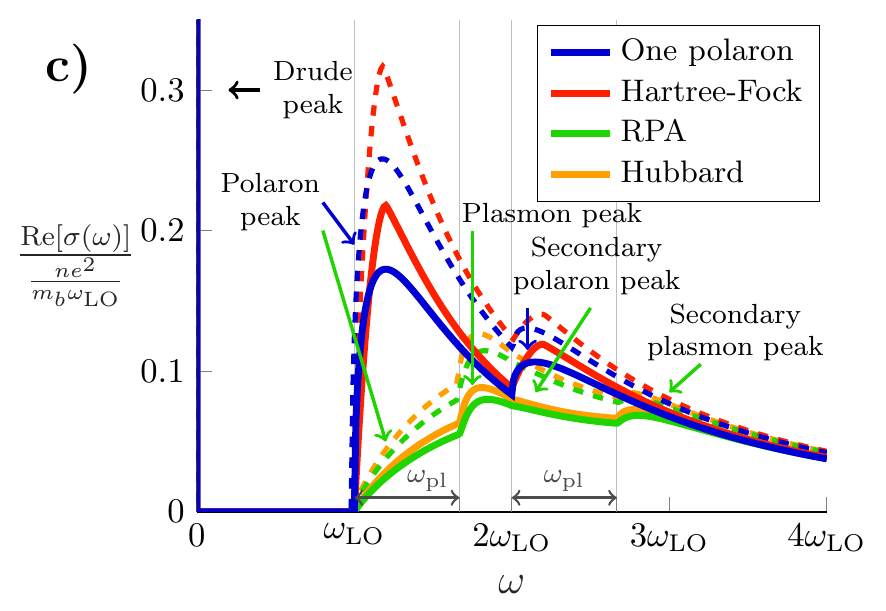}
\includegraphics[width=8.1cm]{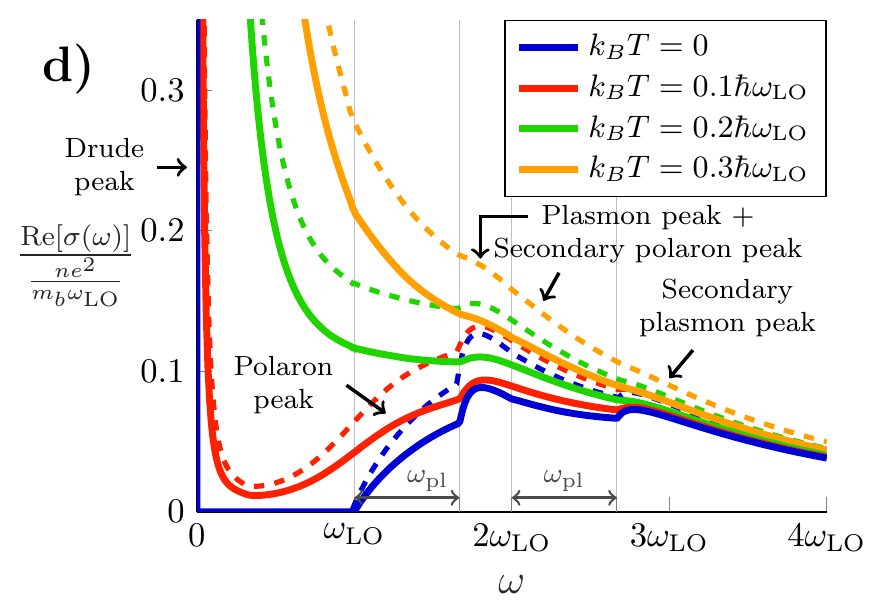}
\caption{\label{fig:CondReal} Optical polaron absorption spectra $\text{Re}[\sigma(\omega)]$ of the anharmonic polaron gas, calculated for several different material parameters. Solid lines represent the results when 3-phonon interaction is excluded ($\mathcal{T_0} = 0$), and dashed lines represent the results including 3-phonon interaction ($\mathcal{T_0} = 0.01$). The origin of the indicated peaks is discussed in the main text. \textbf{a)} Single polaron optical absorption at temperature zero, clearly showing the main polaron peak originating at $\omegaLO$ and a secondary ``anharmonic'' polaron peak originating at $2 \omegaLO$. \textbf{b)-c)} Effect of the structure factor of the polaron gas at temperature zero, calculated using $\mathcal{T}_1 = 0.1$ and \textbf{b}) $r_s = 8$ and \textbf{c)} $r_s = 12$: the existence of plasmons introduces two new peaks. \textbf{d)} Effect of the temperature on the absorption spectrum, calculated using the Hubbard structure factor, $\mathcal{T}_1 = 0.1$ and $r_s = 12$. Note that the Drude peak now has a finite height, but it is still too high to fit in the figure with the other absorption peaks.
\textbf{a)}-\textbf{d)} all use $\alpha = 1$, $\tilde{V}_0 = 0.001$, and $E_{\text{Ry}}/\hbar \omegaLO = 8$.}
\end{figure}

The real part of the optical conductivity contains signatures of polaron formation that can be experimentally measured. Indeed, it is related to the optical absorption coefficient $\Gamma(\omega)$ of a material \cite{Devreese1971}:
\begin{equation}
\Gamma(\omega) = \frac{1}{\varepsilon_{\text{vac}} c \mathsf{N}} \text{Re}[\sigma(\omega)], 
\end{equation}
where $\mathsf{N}$ is the index of refraction of that material. It is known in the literature that a Fr\"ohlich polaron at zero temperature has an absorption peak that appears near $\omegaLO$ (\figref{fig:CondReal}a, blue line), which is usually in the mid-infrared region \cite{Finkenrath1969, Mahan2000}. A many-polaron gas has a second peak that appears near $\omegaLO + \omega_{\text{pl}}$ (see \figref{fig:CondReal}b-c, green and orange lines) due to the formation of plasmons \cite{Tempere2001, Alexandrov2010}. In this section, the effect of the 1-electron-2-phonon interaction on the optical absorption spectrum is investigated. The results are shown in \figref{fig:CondReal}.

The most important result is that the 1-electron-2-phonon interaction leads to an additional absorption peak, which can be seen on the absorption spectra of a single polaron in \figref{fig:CondReal}a. For a single polaron at temperature zero, without 3-phonon terms, the real part of the conductivity actually has an exact expression:
\begin{align}
\text{Re}[\sigma(\omega)] & = \frac{\pi}{2} \frac{n e^2}{m_{\text{eff}}} \delta(\omega) \\
& \hspace{10pt} + \frac{n e^2}{m_b} \frac{2 \alpha}{3} \frac{\omegaLO^{3/2}}{\omega^3} \left( \sqrt{\omega-\omegaLO} \Theta(\omega-\omegaLO) + \frac{2 \mathcal{T}_1^2}{15 \tilde{V}_0} \sqrt{\omega-2 \omegaLO} \Theta(\omega-2 \omegaLO)\right). \label{AnalyticalResult}
\end{align}
This expression has three terms, respectively representing the infinitely sharp Drude peak, the polaron absorption peak (whose expression is well-known in the literature \cite{Mahan2000, Alexandrov2010}), and a new ``secondary polaron peak'' that is due to the new 1-electron-2-phonon interaction process. Each of these peaks is visible in \figref{fig:CondReal}a. The third term only contributes when $\omega > 2 \omegaLO$, which can be understood in the following way. The optical conductivity is zero below $\omega < \omegaLO$ because there are no phonons naturally present to interact with at zero temperature, so an energy of at least $\hbar \omegaLO$ is necessary to create a phonon \cite{Alexandrov2010}. Similarly, two phonons with a total energy of $2 \hbar \omegaLO$ must be created before the 1-electron-2-phonon process of \figref{fig:HamRepresentation}c can contribute to the conductivity, which requires $\hbar \omega > 2 \hbar \omegaLO$.

The secondary polaron peak can serve as an experimental fingerprint for beyond-Fr\"ohlich electron-phonon interaction. Its presence indicates that a 1-electron-2-phonon interaction term like the one in \figref{fig:HamRepresentation}c is not negligible. Additionally, from \eqref{AnalyticalResult} it can be seen that the height of the new peak is proportional to $\mathcal{T}_1^2$. Therefore, a measurement of the height of the secondary polaron peak provides a way to estimate the value of $\mathcal{T}_1$ in a material.

At realistic doping densities, there is never a single polaron, but rather a gas of polarons. \figref{fig:CondReal}b and \figref{fig:CondReal}c show the conductivity of the polaron gas at two different electron densities, using different models for the dynamical structure factor. Firstly, we motivate \emph{a posteriori} that the RPA model \eqref{SFLindhard} for the dynamical structure factor is sufficient to capture the main features of the optical absorption spectra, by noting that in \figref{fig:CondReal}b-c the RPA model captures the same qualitative effects as the Hubbard model: adding the exchange and correlation effects of the electron gas up to lowest order using the Hubbard model does not fundamentally change the absorption spectra. On the contrary, at the densities used in \figref{fig:CondReal}, the single polaron and Hartree-Fock models fail to account for plasmon formation in the polaron gas \cite{Tempere2001} and therefore do not capture the fundamental features of the absorption spectrum. Because of the formation of plasmons, the integral of the structure factor over all momenta (shown in \figref{fig:StructureHubbard}) has a kink at the plasma frequency $\omega_{\text{pl}}$, which leads to an additional peak in the optical absorption spectra of  \figref{fig:CondReal}b-c at $\omegaLO + \omega_{\text{pl}}$. The interplay of both 1-electron-2-phonon interaction and plasmons can finally lead to a secondary plasmon peak. 
Therefore, up to four distinct features may appear in the absorption spectrum: the polaron absorption peak which starts at $\omegaLO$, the plasmon peak at $\omegaLO + \omega_{\text{pl}}$, the secondary polaron absorption peak at $2\omegaLO$, and a secondary plasmon peak at $2\omegaLO + \omega_{\text{pl}}$. Note that, depending on the relative strength of the couplings, some peaks may appear as shoulders or as kinks in the spectrum. Additionally, some of these features may overlap: for example, in \figref{fig:CondReal}c only three peaks are visible, because the plasmon peak and the secondary polaron peak cannot be distinguished from each other. 

\figref{fig:CondReal}d shows the results calculated at finite temperatures. In this case, the relaxation time $\tau_{\text{eff}}$ in expression \eqref{condSigma0} becomes finite, which causes broadening and smearing of the different peaks. Perhaps most notably, the Drude peak broadens from a Dirac delta function to a peak with finite height and width. The Drude peak is much taller than the other absorption peaks and therefore quickly dominates the whole absorption spectrum. In order to discern each of the absorption peaks, the spectrum should be measured at sufficiently low temperatures ($k_B T \lesssim 0.1 \hbar \omegaLO$).

The dashed lines of figure \figref{fig:CondReal} show the result when the 3-phonon interaction is included. The results are qualitatively the same: the 3-phonon interaction only changes the height of the peaks, and moves the locations of the peaks very slightly. Overall, the 3-phonon interaction seems to be most impactful when there is also 1-electron-2-phonon interaction.

\subsection{Qualitative prediction for the DC resistivity}
\begin{figure}
\centering
\includegraphics[width=8.6cm]{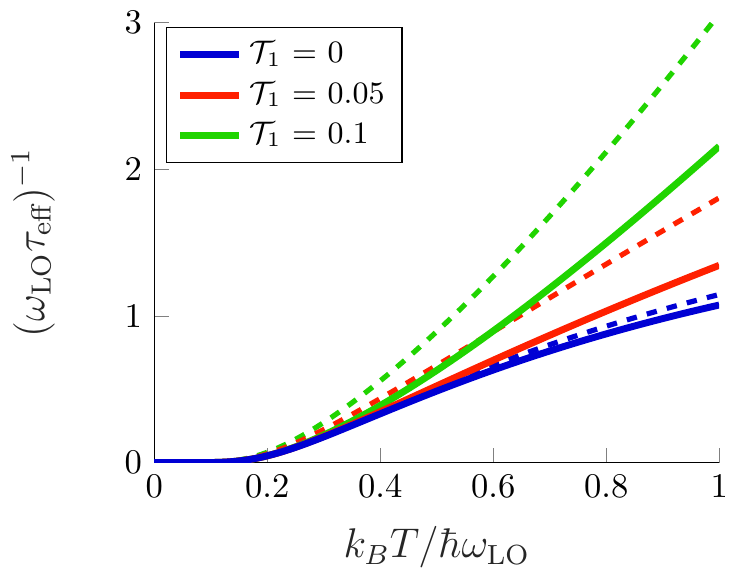}
\caption{\label{fig:ScatteringRateHubbard} The scattering rate $\tau_{\text{eff}}^{-1}$ of the electron-phonon collisions as a function of temperature. At zero temperature, no phonons are present and $\tau_{\text{eff}}^{-1}=0$. Figure made with $\alpha = 1$, $\tilde{V}_0 = 0.001$, $r_s = 12$, $E_{\text{Ry}}/\hbar \omegaLO = 8$, the Hubbard structure factor, and $\mathcal{T}_0 = 0$ (solid lines) or $\mathcal{T}_0 = 0.01$ (dashed lines).}
\end{figure}
The value of the electron-phonon scattering rate $\tau_{\text{eff}}^{-1}$ is shown in \figref{fig:ScatteringRateHubbard}. This scattering rate is also directly proportional to the DC resistivity of the polaron gas $\rho(0) = \frac{m_{\text{eff}}}{ne^2 \tau_{\text{eff}}}$, and therefore also represents the inverse of the height of the Drude peak in \figref{fig:CondReal}d. The scattering rate is thermally activated, and remains almost unchanged by the 1-electron-2-phonon interaction at low temperatures. This can be understood by noting that the 1-electron-2-phonon process requires two phonons to be present in the material: at low temperatures, this is much more unlikely than finding just a single phonon. In the low-temperature limit, using the one-polaron structure factor, and ignoring the 3-phonon interaction, the scattering rate can be calculated using \eqref{tauEffDef}:
\begin{equation} \label{ResistivityOnePolaron}
\tau_{\text{eff}}^{-1} \approx \omegaLO \frac{4\alpha}{3} \frac{\hbar \omegaLO}{k_B T} \left( e^{-\frac{\hbar \omegaLO}{k_B T}} + \frac{2 \sqrt{2} \mathcal{T}_1^2}{15 \tilde{V}_0} e^{-\frac{2\hbar \omegaLO}{k_B T}} \right),
\end{equation}
which highlights the finding that the 1-electron-2-phonon interaction only starts significantly contributing to the scattering rate at $k_B T \approx 2 \hbar \omegaLO$, while the Fr\"ohlich interaction already contributes at $k_B T \approx \hbar \omegaLO$. The result \eqref{ResistivityOnePolaron} reduces to the well-known result in the literature if $\mathcal{T}_1 = 0$ \cite{Alexandrov2010, Mahan2000, Feynman1962}.

There is a known issue with the calculation of the collision rate $\tau_{\text{eff}}^{-1}$ from the Kubo formula. The electron-phonon collision rate can also be calculated from the Boltzmann equation \cite{Kadanoff1963}. Comparing the result for $\mathcal{T}_1 = 0$ in \figref{fig:ScatteringRateHubbard} with the result in \cite{Kadanoff1963} finds that \figref{fig:ScatteringRateHubbard} is wrong by a factor $3 k_B T/2\hbar \omegaLO$ \cite{Peeters1983_kT}. The difference can be interpreted as an incorrect exchange of limits \cite{Peeters1983_kT}: using the Boltzmann equation correctly calculates $\underset{\alpha \rightarrow 0}{\lim} \underset{\omega \rightarrow 0}{\lim} \sigma(\omega)$, whereas using the Kubo formula calculates the limit $\underset{\omega \rightarrow 0}{\lim} \underset{\alpha \rightarrow 0}{\lim} \sigma(\omega)$ which in this case leads to a different result. The other curves in \figref{fig:ScatteringRateHubbard} are presumably also incorrect by a similar factor, which might be different when 1-electron-2-phonon interaction is included. Therefore, the results in \figref{fig:ScatteringRateHubbard} should be seen as a qualitative comparison with the Fr\"ohlich result, rather than quantitative predictions.

\section{Conclusions and outlook} \label{sec:Conclusion}
In this article, we have calculated the optical conductivity and its related quantities for an anharmonic large many-polaron gas, most notably including the 1-electron-2-phonon interaction of \figref{fig:HamRepresentation}c. The Hamiltonian \eqref{HamTot}-\eqref{HamElPh} from \cite{Houtput2021} is an extension of the Fr\"ohlich Hamiltonian with analytical expressions for the interaction strengths. In the low- and high-frequency limits, the conductivity can be written as a Drude conductivity. In the low-frequency limit, the carrier mass is equal to the anharmonic polaron mass \cite{Houtput2021}, which verifies the intuition that anharmonic polarons play the role of charge carriers in this regime.

The optical absorption spectrum is proportional to the real part of the optical conductivity. A gas of Fr\"ohlich polarons has two peaks: a polaron peak at $\omegaLO$, and a polaron-plasmon peak at $\omegaLO + \omega_{\text{pl}}$ \cite{Tempere2001}. In this paper, we have shown that a gas of anharmonic polarons has two more peaks: an anharmonic polaron peak at $2 \omegaLO$, and an anharmonic polaron-plasmon peak at $2\omegaLO + \omega_{\text{pl}}$.

The method used in this paper is strongly based on the method first proposed in \cite{Tempere2001} to calculate the optical conductivity of a gas of Fr\"ohlich polarons, starting from the Kubo formula. One difference with the method in \cite{Tempere2001} is that we use the memory function formalism to introduce the spectral function $M(\mathbf{k},\omega)$ of the phonons, which is then calculated using the Matsubara-Green's formalism. This has the advantage that the optical conductivity can also be calculated at finite temperatures. Additionally, the method proposed in this article works for any Hamiltonian of the following form:
\begin{equation} \label{GeneralHamiltonian}
\hat{H} = \hat{H}_{\text{el}} + \hat{H}_{\text{ph}} + \sum_{\mathbf{q} \neq \mathbf{0}} \hat{\mathcal{F}}_{\mathbf{q}} \hat{\rho}_{-\mathbf{q}},
\end{equation}
where the operators $\hat{H}_{\text{ph}}$ and $\hat{\mathcal{F}}_{\mathbf{q}}$ depend on one or more phonon operators, and the dynamical structure factor $S(\mathbf{k},\omega)$ must be calculated with respect to the general electron Hamiltonian $\hat{H}_{\text{el}}$. With the choice \eqref{FAuxDef} for $\hat{\mathcal{F}}_{\mathbf{q}}$ and the choice \eqref{HamPh} for $\hat{H}_{\text{ph}}$, the above Hamiltonian reduces to \eqref{HamTot}-\eqref{HamElPh}. Regardless, many other electron-phonon Hamiltonians can be written in the form \eqref{GeneralHamiltonian}. One example is the impurity-boson Hamiltonian in ultracold gases, written in the Bogolioubov approximation and including the 1-impurity-2-boson interaction \cite{Rath2013, Ichmoukhamedov2019}. For such Hamiltonians, the weak-coupling conductivity is still given by expressions \eqref{condMemory}-\eqref{MemoryFullDef}, in combination with \eqref{SigmaReGen}-\eqref{SigmaImGen}. If one can calculate the dynamical structure factor $S(\mathbf{k},\omega)$ and the phonon spectral function $M(\mathbf{k},\omega)$ for the Hamiltonian in question, the conductivity can be calculated using this method.

The anharmonic electron-phonon Hamiltonian \eqref{HamTot}-\eqref{HamElPh} in its current form is quite limited in its application to real materials, since it only applies to cubic materials and only contains interaction to a single phonon branch. Furthermore, the most commonly used materials that satisfy those conditions do not have significant 1-electron-2-phonon interaction, as shown by the low values of $\mathcal{T}_1$ in \tabref{tab:MaterialParameters}. In order to study currently relevant anharmonic materials with electron-phonon interaction, like SrTiO$_3$ \cite{Gastiasoro2020} or high-pressure sulfur hydride \cite{Drozdov2015}, the Hamiltonian \eqref{HamTot}-\eqref{HamElPh} must first be generalized to include multiple phonon branches and noncubic point groups. Fortunately, such a Hamiltonian would be of the form \eqref{GeneralHamiltonian}. Therefore, the theory presented in \secref{sec:Theory} of this article can still be applied. For a noncubic material, the conductivity and the memory function $\Sigma_0(\omega)$ will become a $3 \times 3$ matrix, so that equations \eqref{SigmaReGen}-\eqref{SigmaImGen} become:
\begin{align}
\text{Re}\left[\Sigma^{(0)}_{ij}(\omega)\right] = & \frac{2 \omega}{\pi} \mathcal{P} \int_{0}^{+\infty}  \frac{\text{Im}\left[\Sigma^{(0)}_{ij}(\nu)\right]}{\nu^2-\omega^2} 
\tildeff \nu, \\
\text{Im}\left[\Sigma^{(0)}_{ij}(\omega)\right] \approx & -\frac{\pi}{m_b \hbar \omega} \sum_{\mathbf{k}} k_i k_j \int_{-\infty}^{+\infty} \left[1+n_B(\omega')+n_B(\omega-\omega')\right] S(\mathbf{k},\omega-\omega') M(\mathbf{k},\omega') d\omega' .
\end{align}
To use these expressions, only the spectral function $M(\mathbf{k},\omega)$ and perhaps $S(\mathbf{k},\omega)$ would have to be recalculated for the material in question.

An interesting avenue to explore further is the effect of 3-phonon interaction on the optical conductivity of the polaron gas. In section \ref{sec:PhononSpectralFunction}, we motivated why the 3-phonon interaction \eqref{HamPh} used in this article is insufficient because it only contains interactions between LO phonons. With a Hamiltonian that contains all possible 3-phonon processes \cite{Srivastava2019}, it would be possible to properly study the effect of 3-phonon interaction on the optical conductivity. Expressions \eqref{MFromF} and \eqref{F3ph} suggest that only the phonon self energies or the phonon spectral functions are sufficient to calculate $M(\mathbf{k},\omega)$ and study the effect of 3-phonon interaction, as long as no 1-electron-2-phonon interaction is considered.

It is possible to calculate the conductivity of the anharmonic polaron gas described by the Hamiltonian \eqref{HamTot}-\eqref{HamElPh} using several other methods, which would give results in different regimes. For example, to calculate the optical conductivity of one polaron at intermediate or strong electron-phonon coupling, the path integral method of \cite{Feynman1962, Devreese1972} or the diagrammatic Monte Carlo method of \cite{Mishchenko2003} could be generalized. In order to calculate the electron-phonon scattering rate $\tau_{\text{eff}}^{-1}$ and the DC conductivity of the polaron gas, and to verify whether the correct value is still obtained after multiplying with a factor $3 k_B T/2 \hbar \omegaLO$, one could use the Boltzmann transport equation as in \cite{Kadanoff1963}. Both of these are left as potential further research questions.

We propose the anharmonic polaron absorption peak at $2\omegaLO$ as an experimental fingerprint for 1-electron-2-phonon interaction in solids. Since the height of the anharmonic polaron peak is proportional to $\mathcal{T}_1^2$, a measurement of the height of this peak can be used to estimate the relevance of the 1-electron-2-phonon interaction in a material.

\section*{Code availability}
The code that was used to generate \figref{fig:CondImag}-\ref{fig:ScatteringRateHubbard} is publically available online at: \url{https://github.com/MHoutput/AnharmonicPolaronConductivity}.

\acknowledgments

This research was funded by the University Research Fund (BOF) of the University of Antwerp (project ID: 38499). We would like to thank S. Klimin and T. Ichmoukhamedov for many interesting discussions and suggestions on the calculations and the results. We also thank L. Ranalli, C. Verdi, C. Franchini and G. Kresse from the University of Vienna for discussions and especially for their \emph{ab initio} calculation of the anharmonic coefficients $\mathcal{T}_1$ in \tabref{tab:MaterialParameters}.

\bibliography{References}

\end{document}